\documentclass[preprint,review,12pt]{elsarticle}

\usepackage{amssymb}
\usepackage{amsmath}

\usepackage[hyperindex,breaklinks,hidelinks,colorlinks,citecolor=blue]{hyperref}
\usepackage{graphicx,float}
\usepackage{dcolumn}
\usepackage{bm}
\usepackage{afterpage}
\usepackage[T1]{fontenc}
\usepackage{longtable}

\hypersetup{
    colorlinks=true, 
    linkcolor=black, 
    urlcolor=blue    
}

\journal{Chem. Eng. Sci. (10.1016/j.ces.2025.121566)}
\date{January 17, 2025}
\begin{document}

\begin{frontmatter}



\title{Forcing mass transfer approach in multicomponent miscible mixtures using the lattice Boltzmann method}

 \author[PEQ]{Ramon G. C. Lourenço}
 \author[DEQ]{Pedro H. Constantino\corref{cor1}}
 \author[PEQ,DEQ,EPQB]{Frederico W. Tavares}
 
 \affiliation[PEQ]{
             organization={Chemical Engineering Program, PEQ/COPPE},
             addressline={Federal University of Rio de Janeiro, PO Box 68502},
             city={Rio de Janeiro},
             postcode={21941-972},
             state={RJ},
             country={Brazil}}
 \affiliation[DEQ]{
             organization={Chemical Engineering Department (DEQ)},
             addressline={School of Chemistry (EQ), Federal University of Rio de Janeiro, P.O. Box 68542},
             city={Rio de Janeiro},
             postcode={21941-909},
             state={RJ},
             country={Brazil}}
 \affiliation[EPQB]{
             organization={Program in Engineering of Chemical and Biochemical Processes, EPQB},
             addressline={Federal University of Rio de Janeiro, P.O. Box 68542},
             city={Rio de Janeiro},
             postcode={21941-909},
             state={RJ},
             country={Brazil}}
\cortext[cor1]{constantino@eq.ufrj.br (Corresponding author)}

\begin{abstract}
While the lattice Boltzmann method (LBM) has proven robust in areas like general fluid dynamics, heat transfer, and multiphase modeling, its application to mass transfer has been limited. Current modeling strategies often oversimplify the complexities required for accurate and realistic mass transfer simulations in multicomponent miscible mixtures involving external forces. We propose a forcing approach within the explicit velocity-difference LBM framework to address these limitations. Our approach recovers the macroscopic mass conservation equations, the Navier-Stokes equation with external forcing term, and the full Maxwell-Stefan equation for ideal mixtures at low Knudsen numbers. A novel boundary scheme for impermeable solid walls is also suggested to ensure proper mass conservation while effectively managing the spatial interpolations required for multicomponent mixtures with varying molecular masses. We demonstrated the physical consistency and accuracy of the proposed forcing approach through simulations of the ultracentrifuge separation of uranium isotopes and the Loschmidt tube with gravitational effects. Our approach encompasses advanced modeling of species dynamics influenced by force fields, such as those encountered in geological CO$_2$ sequestration in aquifers and oil reservoirs under gravitational fields.
\end{abstract}


\end{frontmatter}



\section{Introduction}\label{sec:introduction}

In recent years, the lattice Boltzmann method (LBM) has significantly expanded into diverse fields such as fluid dynamics, thermal processes, and complex material modeling \cite{Sharma2019}. This growth is attributed to its robust capabilities, including incorporating thermodynamic effects \cite{Ridl&Wagner2018, Li_Xing_Huang2023} and leveraging parallel computing for efficient simulations \cite{Feichtinger_etal2015}. Among the fields where the LBM has found application, mass transfer stands out as a critical area of study. A thorough understanding of its implications can contribute to refining technological systems and developing more effective processes,  including those employed in solid oxide and proton exchange membrane fuel cells \cite{BLESZNOWSKI2022121878,Wang_etal2023}.

The LBM for mass transfer modeling is broadly categorized into single-fluid and multifluid strategies \cite{LuoGirimaji2002}. The single-fluid strategy, which includes passive scalar \cite{Flekkoy1993, Dawson1993} and force models like the pseudopotential approach \cite{Lourenço2022a, Shan&Chen1993}, often simplifies the system. While these models are known for their computational efficiency and ease of implementation \cite{Lourenço2022b,Hosseini2018, Chai2019}, they fall short when applied to complex multicomponent miscible systems, particularly in non-dilute conditions \cite{Lourenço2024}. In contrast, the multifluid strategy provides a more detailed representation by distinguishing the properties of individual species, such as their specific contributions to viscosity and velocity within the mixture \cite{Sirovich1962, Hamel1965, Gorban&Karlin1994}. This strategy comprises three main modeling approaches: (i) the quasi-equilibrium models, (ii) the equilibrium-adapted models, and (iii) the explicit velocity-difference models \cite{Lourenço2024}. 

Quasi-equilibrium models use Boltzmann's H-theorem to offer mathematical advantages \cite{Arcidiacono&Mantzaras_etal2006} and simulate ideal mixtures effectively \cite{Sawant2021}, while the equilibrium-adapted models handle mixtures with varying molecular masses without interpolation, delivering a robust mathematical framework \cite{AsinariMulticomponent2008, Asinari2009}. However, both approaches perform a variable transformation that alters the first-order moment, requiring the solution of local linear systems for species velocities, which can be computationally intensive \cite{Arcidiacono2007, Zudrop2017,Sawant2021}. Explicit velocity-difference (EVD) models provide a structured method for addressing mass transport issues and have advanced significantly over the years, driven by contributions from various research groups \cite{Lourenço2024}. This collaborative development has introduced multiple viewpoints, enriching the model and enhancing its versatility.

Building on the foundational work of Sirovich \cite{Sirovich1962}, the initial developments of the EVD model focused on binary miscible mixtures with species of identical molecular mass \cite{LuoGirimaji2002, LuoGirimaji2003}. The subsequent extension to binary and multicomponent mixtures with varying molecular masses required strategies such as spatial interpolations to effectively manage different thermal velocities within a single grid domain, as well as iterative adjustments for cross-collision parameters \cite{Joshietal2007a, Joshietal2007b, Tong_etal2014}. While some approaches aim to avoid interpolating \cite{Xu&Dang2016}, doing so enhances numerical stability and reduces the required lattice resolution \cite{McCracken2005}. Additionally, Chapman-Enskog (C-E) analysis links the relaxation parameters to macroscopic diffusivities, revealing that the mass and momentum conservation equations and the Maxwell-Stefan (M-S) closure equation for mass transfer emerge for low Knudsen numbers \cite{Tong_etal2014}. These efforts have led to several applications in the literature, including dry reforming of methane \cite{Lin_etal2021a}, mass transport combined with turbulence dynamics \cite{Li_etal2020, WangZhaoNie2019}, porous media \cite{Ma&Chen2015}, and solid oxide fuel cells \cite{Xu&Dang2016, Dang&Xu2016, XuDangBai2014}. Readers are encouraged to consult the recent review in Ref. \cite{Lourenço2024} for a more in-depth exploration of mass transfer modeling with LBM.

Despite these advancements, the application of the EVD model to complex technological problems still poses unresolved issues. No current approach of the EVD-LBM is capable of addressing species dynamics with forcing effects, which is essential for realistic and accurate simulations, including gravitational effects in multicomponent mixtures such as reservoir systems \cite{Nikpoor_etal2016}, electrostatic interactions in applications like electrolytic cells \cite{Ryan&Mukherjee_2019}, and species-surface interactions observed in porous materials \cite{Santos_etal2024}. To the best of our knowledge, this is the first comprehensive attempt to extend the EVD-LBM approach for modeling multicomponent dynamics with external forces in miscible mixtures, ensuring that the M-S equation is fully recovered. To enhance the implementation of the proposed forcing approach, we present a novel halfway bounce-back scheme for impermeable solid walls that guarantees mass conservation at the boundaries, even when utilizing the demanded spatial interpolations. The investigation that will be provided can also facilitate future integration of multiphase modeling and non-ideal mixture behavior, expanding the applicability.

Hence, we propose a forcing approach in the EVD model, which, through the revisited C-E analysis, recovers the Navier-Stokes equation (NSE) with the forcing term and the full M-S equation for ideal mixtures at low Knudsen numbers. This work is organized as follows. Section \ref{sec:theory} presents the state-of-the-art in EVD models. Section \ref{sec:forcing_term} introduces the forcing approach within the model and conducts a C-E analysis. Section \ref{sec:simulations} validates the approach and discusses the results, while Section \ref{sec:conclusion} summarizes the key contributions and future directions of this work.

\section{The explicit velocity-difference (EVD) model}\label{sec:theory}

The EVD model is categorized as a split collision scheme \cite{BGK_split_1954, Gross&Krook_split_1956}. In this model, a self-collision term $\Omega^{ii}$ captures the collisions between particles of the same species, while a cross-collision term $\Omega^{ij}$ accounts for collisions between particles of different species. As a result, the discrete form of the Boltzmann transport equation (BTE), namely the lattice Boltzmann equation (LBE), arises,
\begin{equation} \label{eq:LBE_Sirovich}
    f^i_\alpha (\mathbf{x} + \mathbf{e}^i_\alpha \delta t, t + \delta t) = f^i_\alpha (\mathbf{x}, t) + \left[ \Omega^{ii}_\alpha(\mathbf{x}, t) + \sum_{j \neq i}^N \Omega^{ij}_\alpha(\mathbf{x}, t) \right] \delta t \; ,
\end{equation}
where $i$ and $j$ represent the species, $f^i_\alpha(\mathbf{x}, t)$ denotes the probability distribution function of finding a particle $i$ at a given time $t$ and position $\mathbf{x}$ in the discrete velocity space, and $\mathbf{e}_\alpha^i$ is the dimensionless discrete velocity of the species $i$ in the lattice direction $e_\alpha$.

The self-collision contribution usually assumes the standard Bhatnagar-Gross-Krook-like approximation,
\begin{equation} \label{eq:BGK_self_collision}
    \Omega^{ii}_\alpha = -\frac{1}{\tau_i} \left(f^i_\alpha - f^{i(0)}_\alpha \right) \; ,
\end{equation}
and the cross-collision contribution is expressed by expanding $f^i_\alpha$ around its equilibrium value $f^{i(0)}_\alpha$ under isothermal conditions \cite{LuoGirimaji2002, LuoGirimaji2003}, which yields
\begin{equation} \label{eq:Cross_collision_Sirovich}
    \Omega^{ij}_\alpha = -\frac{1}{\tau_{ij}} \left( \frac{\rho_j}{\rho} \right) \frac{f^{i(eq)}_\alpha}{c_{s,i}^2}  (\mathbf{e}^i_\alpha - \mathbf{u}) \cdot \left(\mathbf{u}_i^{eq} - \mathbf{u}_j^{eq} \right) \; ,
\end{equation}
where $\tau_i$ and $\tau_{ij}$ are the relaxation time parameters, $c_{s,i}$ represents the speed of sound for species $i$, $\rho_j$ and $\rho$ denote the species and mixture densities, $\mathbf{u}_i^{eq}$ is the species equilibrium velocity, and $\mathbf{u}$ is the mixture velocity. By performing a second-order Hermite expansion, the equilibrium distribution is expressed as
\begin{equation} \label{eq:feqi0}
    f^{i(0)}_\alpha (\mathbf{x},t) = \left[1+ \frac{1}{c_{s,i}^2} (\mathbf{e}^i_\alpha - \mathbf{u}) \cdot \left(\mathbf{u}_i^{eq} - \mathbf{u} \right) \right] f^{i(eq)}_\alpha(\mathbf{x},t) \; ,
\end{equation}
\begin{equation}
    f^{i(eq)}_\alpha (\mathbf{x},t) = \omega_\alpha \rho_i(\mathbf{x},t) \left[1 + \frac{\mathbf{e}^i_\alpha \cdot \mathbf{u}}{c_{s,i}^2} + \frac{(\mathbf{e}^i_\alpha \cdot \mathbf{u})^2}{2c_{s,i}^4} - \frac{\mathbf{u}^2}{2c_{s,i}^2} \right] \; ,
\end{equation}
where $\omega_\alpha$ are weights.

As a result of this multifluid approach, species with varying molecular masses $M_i$, and consequently different thermal speeds at a given temperature, have distinct sets of discrete velocities \cite{McCracken2005}. This feature can be incorporated into the method using the different lattice speeds (DLS) scheme, where the dimensionless thermal speed $c_i$ of the lightest species ($i=1$) is primarily set, often as $c_1=1$. Since the thermal speed and sound speed are related to the molecular masses of the species, $c_i$ and $c_{s,i}$ are determined accordingly,
\begin{equation} \label{eq:speed_DLS}
    \frac{c_i}{c_1} = \frac{c_{s,i}}{c_{s,1}} = \sqrt{\frac{M_1}{M_i}} \; .
\end{equation}
As the dimensionless discrete velocities depend on the thermal speeds, $\mathbf{e}_\alpha^i$ is expressed for the D2Q9 lattice arrangement by \cite{McCracken2005} 
\begin{equation} \label{eq:discrete_velocity_DLS}
    \mathbf{e}^i_\alpha =\left\{ \begin{matrix}  \multicolumn{1}{l}{0 \;, \;\alpha=0} \\ \multicolumn{1}{l}{c_i \left[ \cos \left( \frac{(\alpha-1)\pi}{2} \right),\sin \left( \frac{(\alpha-1)\pi}{2} \right)  \right] \; , \; \alpha = 1,2,3,4} \\ \multicolumn{1}{l}{c_i\sqrt{2}\left[ \cos \left( \frac{(\alpha-5)\pi}{2} + \frac{\pi}{4} \right),\sin \left( \frac{(\alpha-5)\pi}{2} + \frac{\pi}{4} \right)  \right] \; , \; \alpha = 5,6,7,8}  \end{matrix} \right. \; .
\end{equation}

\begin{figure}[h]
    	\centering
    	\includegraphics[width=0.5\linewidth]{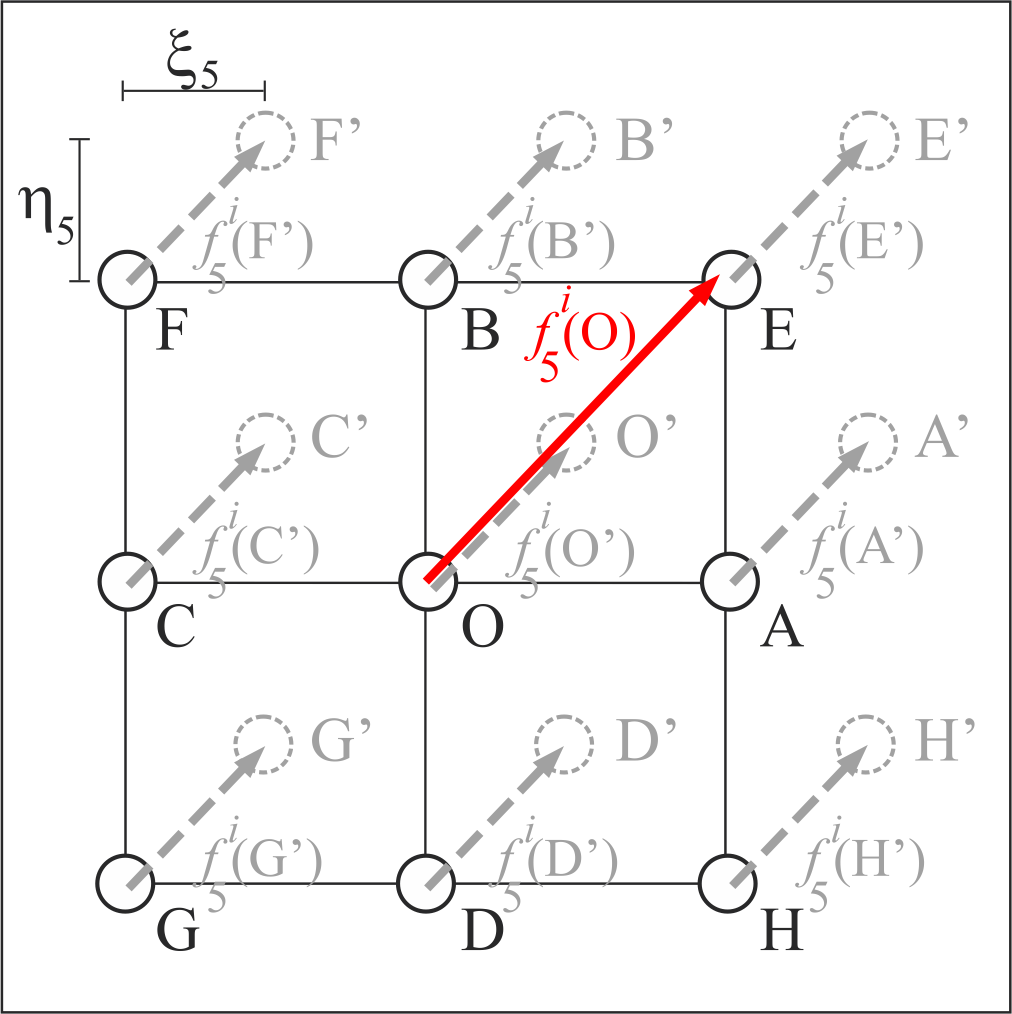}
    	\caption{The second-order spatial interpolation scheme is demonstrated for $\alpha = 5$. Solid black circles, denoted by pure letters, mark the discrete lattice points, while dashed light-grey circles, indicated by single-primed letters, represent the off-lattice positions to which species $i \neq 1$ have moved. $f^i_\alpha$ is calculated using spatial interpolation at each node, drawing on the values of $f^i_\alpha(O')$, ..., $f^i_\alpha(H')$, which are known from the previous streaming step.}
    	\label{fig:Second_order_interpolation}
    \end{figure}

A common interpretation of the DLS scheme is that species travel different distances within the same time interval \cite{McCracken2005}. Compared to the reference species $i=1$, which travels a distance $\delta x_1$, any other heavier species cover $\delta x_i \leq \delta x_1$. This implies that spatial interpolations are necessary to align the distribution functions with the correct lattice locations during species propagation. One such interpolation scheme employs a second-order approach to ensure accurate positioning of the distribution functions \cite{Tong_etal2014},
\begin{equation}  
    \begin{aligned} \label{eq:Second_order_spatial_interpolation}
    f_\alpha^i(O) = (1-\eta_\alpha^2)(1-\xi_\alpha^2)f_\alpha^i(O') + \frac{\xi_\alpha(1-\eta_\alpha^2)(1+\xi_\alpha)}{2}f_\alpha^i(C') \\ + \frac{\eta_\alpha (1+\eta_\alpha) (1-\xi^2_\alpha)}{2} f_\alpha^i(D') + \frac{\eta_\alpha \xi_\alpha (1+\eta_\alpha) (1+\xi_\alpha)}{4} f_\alpha^i(G') \\ - \frac{\xi_\alpha (1-\xi_\alpha) (1-\eta^2_\alpha)}{2} f_\alpha^i(A') - \frac{\eta_\alpha (1-\eta_\alpha) (1-\xi^2_\alpha)}{2} f_\alpha^i(B') \\ - \frac{\eta_\alpha \xi_\alpha (1+\xi_\alpha) (1-\eta_\alpha)}{4} f_\alpha^i(F') - \frac{\eta_\alpha \xi_\alpha (1-\xi_\alpha) (1+\eta_\alpha)}{4} f_\alpha^i(H') \\ + \frac{\eta_\alpha \xi_\alpha (1-\xi_\alpha) (1-\eta_\alpha)}{4} f_\alpha^i(E') \; ,
    \end{aligned}
\end{equation}
where $\xi_\alpha = e^i_\alpha |_x \delta t $, $\eta_\alpha = e^i_\alpha |_y \delta t$, and $\mathbf{e}^i_\alpha = (e^i_\alpha |_x, \; e^i_\alpha |_y)$. As shown in Fig. \ref{fig:Second_order_interpolation}, a plain letter represents the node position corresponding to the lattice domain, whereas a letter with a single prime denotes the virtual adjacent node following the streaming of species $i \neq 1$. Hence, the distribution functions $f_\alpha^i(O')$ obtained during the propagation step are corrected at each lattice node using Eq. (\ref{eq:Second_order_spatial_interpolation}).

Once they have been processed and boundary conditions implemented, the species density and species equilibrium velocity are derived from the zeroth- and first-order moments of $f^i_\alpha$,
\begin{equation} \label{eq:zeroth_order_moment}
    \rho_i (\mathbf{x},t) = \sum_\alpha f^i_\alpha (\mathbf{x},t) \qquad \; ,
\end{equation}
\begin{equation} \label{eq:first_order_moment}
    \mathbf{u}^{eq}_i (\mathbf{x},t) = \frac{1}{\rho_i (\mathbf{x},t)} \sum_\alpha \mathbf{e}_\alpha^i f^i_\alpha (\mathbf{x},t) \quad  \; ,
\end{equation}
which can then be used to compute the mixture density and velocity,
\begin{equation} \label{eq:Sirovich_mixture_density}
    \rho (\mathbf{x},t) = \sum_i \rho_i (\mathbf{x},t) \; ,
\end{equation}
\begin{equation} \label{eq:Sirovich_mixture_velocity}
    \mathbf{u} (\mathbf{x},t) = \sum_i w_i (\mathbf{x},t) \mathbf{u}_i^{eq} (\mathbf{x},t) \; ,
\end{equation}
where $w_i = \rho_i/\rho$ is the mass fraction of species $i$.

This methodology effectively captures diffusive and advective mass transfer in miscible multicomponent mixtures, even for species with varying molecular masses. However, when external forces such as gravitational, centrifugal, electrostatic, or interaction forces become significant, the existing approach proves insufficient and requires further enhancement. In light of this, we incorporate the well-established forcing approach from the LBM literature \cite{Li_Luo_Li_2012} and propose a novel analysis arising from its application, as described in Section \ref{sec:forcing_term}.

\section{A forcing approach within the EVD model}\label{sec:forcing_term}

In this section, we present a novel perspective derived from analyzing the forcing approach within the EVD model. The LBE, incorporating the classic forcing term and the collision models advanced by \citet{Sirovich1962} and \citet{LuoGirimaji2002,LuoGirimaji2003}, is investigated using the C-E expansion. The macroscopic governing equations are derived after computing the associated moment equations. We show that the forcing contribution to species dynamics consistently emerges in the momentum conservation and M-S equations. Additionally, the relationships between the relaxation parameters and the diffusivities are demonstrated, thereby clarifying the role of the kinetic parameters. As a starting point, the LBE with the standard forcing term $S_\alpha^i$ reads \cite{Li_Luo_Li_2012}
\begin{equation} \label{eq:LBE_forcing}
    f^i_\alpha (\mathbf{x} + \mathbf{e}^i_\alpha \delta t, t + \delta t) = f^i_\alpha (\mathbf{x}, t) + \left[ \Omega_\alpha^i(\mathbf{x}, t)  + \left( 1 - \frac{\delta t}{2 \tau_i}\right) S_\alpha^i(\mathbf{x}, t) \right] \delta t \; ,
\end{equation}
where $\Omega^{i}_\alpha(\mathbf{x}, t) \equiv \Omega^{ii}_\alpha(\mathbf{x}, t) + \sum_{j \neq i}^N \Omega^{ij}_\alpha(\mathbf{x}, t)$.
Expanding the first term using the Taylor series, rearranging the equation, and omitting the notation  $(\mathbf{x}, t)$ for conciseness,
\begin{equation}
\label{eq:CE_EQ6}
    \begin{aligned}
    \Omega_\alpha^i + \left( 1 - \frac{\delta t}{2 \tau_i}\right) S_\alpha^i = \left(\partial_t + \mathbf{e}^i_\alpha \cdot \mathbf{\nabla} \right) f^i_\alpha \\ + \left[\frac{1}{2} \left(\mathbf{e}^i_\alpha \mathbf{e}^i_\alpha : \mathbf{\nabla} \mathbf{\nabla} \right) + \mathbf{e}^i_\alpha \cdot \partial_t \mathbf{\nabla} + \frac{1}{2} \partial_t^2\right] f_\alpha^i \delta t \; .
    \end{aligned}
\end{equation}

By introducing a small parameter $\varepsilon$ with the same order of the Knudsen number, the time and spatial derivatives, distribution function, and collision and forcing terms are asymptotically expanded as
\begin{equation} \label{eq:CE_time_derivative}
    \partial_t = \varepsilon \partial_t^{(1)} + \varepsilon^2 \partial_t^{(2)} + \mathcal{O}(\varepsilon^3)\; ,
\end{equation}
\begin{equation} \label{eq:CE_spatial_derivative}
    \nabla = \varepsilon \nabla^{(1)} + \mathcal{O}(\varepsilon^2)\; ,
\end{equation}
\begin{equation} \label{eq:CE_distribution_function}
    f_\alpha^{i} = f_\alpha^{i,(0)} + \varepsilon f_\alpha^{i,(1)} + \varepsilon^2 f_\alpha^{i,(2)} + \mathcal{O}(\varepsilon^3)\; ,
\end{equation}
\begin{equation} \label{eq:CE_self_collision_term}
    \Omega_\alpha^{ii} = \Omega_\alpha^{ii,(0)}+ \varepsilon \Omega_\alpha^{ii,(1)} + \varepsilon^2 \Omega_\alpha^{ii,(2)} + \mathcal{O}(\varepsilon^3)\; ,
\end{equation}
\begin{equation} \label{eq:CE_cross_collision_term}
    \Omega_\alpha^{ij} = \varepsilon \Omega_\alpha^{ij,(1)} + \mathcal{O}(\varepsilon^2)\; ,
\end{equation}
\begin{equation} \label{eq:CE_forcing_term}
    S_\alpha^{i} = \varepsilon S_\alpha^{i,(1)} + \mathcal{O}(\varepsilon^2)\; .
\end{equation}

Substituting Eqs. (\ref{eq:CE_time_derivative})--(\ref{eq:CE_forcing_term}) into Eq. (\ref{eq:CE_EQ6}), neglecting terms of order higher than $\mathcal{O}(\varepsilon^2)$, defining $D_\alpha^{(1)} \equiv \partial_t^{(1)} + \mathbf{e}_\alpha^i \cdot \nabla^{(1)}$, and matching terms of the same order, the following three equations arise,
\begin{equation} \label{eq:CE_self_coll_zero}
    \Omega_\alpha^{ii,(0)} = 0\; ,
\end{equation}
\begin{equation} \label{eq:CE_EQ18}
    \Omega_\alpha^{ii,(1)} + \sum_{j \neq i}^N \Omega^{ij,(1)}_\alpha + \left( 1 - \frac{\delta t}{2 \tau_i}\right) S_\alpha^{i,(1)} =  D_\alpha^{(1)} f^{i,(0)}_\alpha \; ,
\end{equation}
\begin{equation} \label{eq:CE_EQ19}
    \Omega_\alpha^{ii,(2)} =\left[\partial_t^{(2)} + \frac{\delta t}{2} \left(D_\alpha^{(1)} \right)^2 \right] f^{i,(0)}_\alpha + D_\alpha^{(1)} f^{i,(1)}_\alpha \; .
\end{equation}

The substitution of Eqs. (\ref{eq:CE_distribution_function}) and (\ref{eq:CE_self_collision_term}) into the self-collision model reveals that $f^{i,(0)}_\alpha$ is actually determined to ensure that Eq. (\ref{eq:CE_self_coll_zero}) is true,
\begin{equation}
    f^{i,(0)}_\alpha = \left[1+ \frac{1}{c_{s,i}^2} (\mathbf{e}^i_\alpha - \mathbf{u}) \cdot \left(\mathbf{u}_i^{eq} - \mathbf{u} \right) \right] f^{i(eq)}_\alpha \; .
\end{equation}
Also after combining the collision models and Eqs. (\ref{eq:CE_distribution_function})--(\ref{eq:CE_cross_collision_term}) with Eqs. (\ref{eq:CE_EQ18}) and (\ref{eq:CE_EQ19}), two pivotal equations of the C-E analysis emerge,
\begin{equation} \label{eq:CE_EQ24}
    -\frac{1}{\tau_i} f_\alpha^{i,(1)} -\frac{f_\alpha^{i(eq)}}{\varepsilon \rho c_{s,i}^2} (\mathbf{e}^i_\alpha - \mathbf{u}) \cdot \sum^N_{j\neq i}\frac{\rho_j}{\tau_{ij}} \left(\mathbf{u}_i^{eq} - \mathbf{u}_j^{eq} \right) + \left( 1 - \frac{\delta t}{2 \tau_i}\right) S_\alpha^{i,(1)} =  D_\alpha^{(1)} f^{i,(0)}_\alpha \; ,
\end{equation}
\begin{equation} \label{eq:CE_EQ27}
    \begin{aligned}
     -\frac{1}{\tau_i} f_\alpha^{i,(2)} = \partial_t^{(2)} f^{i,(0)}_\alpha + \left(1 - \frac{\delta t}{2 \tau_i} \right) D_\alpha^{(1)} \left[f_\alpha^{i,(1)} + \frac{\delta t}{2} S_\alpha^{i,(1)} \right. \\ \left. - \frac{ \delta t \tau_i}{ \left(2 \tau_i - \delta t \right)} \frac{ f_\alpha^{i(eq)}}{\varepsilon \rho c_{s,i}^2} (\mathbf{e}^i_\alpha - \mathbf{u}) \cdot \sum^N_{j\neq i}\frac{\rho_j}{\tau_{ij}} \left(\mathbf{u}_i^{eq} - \mathbf{u}_j^{eq} \right) \right] \; .
    \end{aligned}
\end{equation}

The moments of the previous equilibrium functions, achieved by regarding the isotropy conditions established during the discretization of the BTE \cite{Sbragaglia_etal2007}, are presented in the supplemental material. However, given the inclusion of external forces $\mathbf{F}_i$, it becomes necessary to assess the moments pertaining to the non-equilibrium contributions and the forcing term as well,
\begin{equation} \label{eq:CE_EQ48}
    \sum_\alpha f_\alpha^{i,\text{(neq)}} = - \frac{\delta t}{2} \sum_\alpha S_\alpha^i = - \frac{\delta t}{2} \varepsilon \sum_\alpha S_\alpha^{i,(1)} \; ,
\end{equation}
\begin{equation} \label{eq:CE_EQ49}
    \sum_\alpha \mathbf{e}_\alpha^i f_\alpha^{i,\text{(neq)}} = -\frac{\delta t}{2} \sum_\alpha \mathbf{e}_\alpha^i S_\alpha^i = - \frac{\delta t}{2} \varepsilon \sum_\alpha \mathbf{e}_\alpha^i S_\alpha^{i,(1)} \; ,
\end{equation}
\begin{equation} \label{eq:CE_EQ52}
    \sum_\alpha S_\alpha^i = \sum_\alpha S_\alpha^{i,(1)} = 0 \; ,
\end{equation}
\begin{equation} \label{eq:CE_EQ54}
    \sum_\alpha \mathbf{e}_\alpha^i S_\alpha^i = \varepsilon \sum_\alpha \mathbf{e}_\alpha^i S_\alpha^{i,(1)} = \mathbf{F}_i \; .
\end{equation}

Combining Eqs. (\ref{eq:CE_distribution_function}), (\ref{eq:CE_EQ52}), and (\ref{eq:CE_EQ54}) with Eqs. (\ref{eq:CE_EQ48}) and (\ref{eq:CE_EQ49}), the moments of the non-equilibrium contributions become
\begin{equation} 
    \sum_\alpha f_\alpha^{i,(j)} = 0 \; , \qquad j \geq 1 \; ,
\end{equation}
\begin{equation} 
    \sum_\alpha \mathbf{e}_\alpha^i f_\alpha^{i,(1)} = - \frac{\delta t}{2 \varepsilon} \mathbf{F}_i \; ,
\end{equation}
\begin{equation} 
    \sum_\alpha \mathbf{e}_\alpha^i f_\alpha^{i,(2)} = 0 \; ,
\end{equation}
which means that the mixture density remains calculated by Eq. (\ref{eq:zeroth_order_moment}). However, the equilibrium velocity must be determined by
\begin{equation} \label{eq:first_order_moment_force}
    \mathbf{u}^{eq}_i = \frac{1}{\rho_i } \sum_\alpha \mathbf{e}_\alpha^i f^i_\alpha  + \frac{\mathbf{F}_i \delta t }{2 \rho_i} 
\end{equation}
instead of Eq. (\ref{eq:first_order_moment}).

Under these conditions, and neglecting terms of order higher than $\mathcal{O}(\varepsilon^2)$ or $\mathcal{O}(Ma^2)$, where $Ma$ is the Mach number, the analysis of the zeroth and first moments of Eqs. (\ref{eq:CE_EQ24}) and (\ref{eq:CE_EQ27}) leads to
\begin{equation} \label{eq:CE_EQ59}
    \partial_t^{(1)} \rho_i + \nabla^{(1)} \cdot \rho_i \mathbf{u}_i^{eq} = 0 \; ,
\end{equation}
\begin{equation} \label{eq:CE_EQ67}
\begin{aligned}
    \partial_t^{(1)} \rho_i \mathbf{u}_i^{eq} + \nabla^{(1)} \rho_i c_{s,i}^2 + \nabla^{(1)} \cdot \left[\rho_i \left(\mathbf{u}_i^{eq} \mathbf{u} + \mathbf{u} \mathbf{u}_i^{eq} - \mathbf{u} \mathbf{u} \right) \right] = \frac{1}{\varepsilon} \left(\mathbf{F}_i - \mathbf{\Theta}_i\right) \; ,
\end{aligned}
\end{equation}
\begin{equation} \label{eq:CE_EQ70}
    \partial_t^{(2)} \rho_i - \frac{\delta t}{2 \varepsilon} \nabla^{(1)} \cdot \mathbf{\Theta}_i = 0 \; ,
\end{equation}
\begin{equation} \label{eq:CE_EQ75}
\begin{aligned}
    \partial_t^{(2)} \rho_i \mathbf{u}_i^{eq} = \left(1 - \frac{\delta t}{2 \tau_i} \right) c_{s,i}^2 \nabla^{(1)}  \cdot \tau_i \rho \left[ \varepsilon \nabla^{(1)} w_i \mathbf{u}_i^{eq} + \left( \varepsilon \nabla^{(1)} w_i \mathbf{u}_i^{eq} \right)^\text{T} \right] \\ + \left( 1 - \frac{\delta t}{2 \tau_i} \right) \nabla^{(1)} \cdot \left[ \frac{\tau_i}{\varepsilon} \left( \mathbf{F}_i \mathbf{u} + \mathbf{u} \mathbf{F}_i \right) \right] - \left(1 - \frac{\delta t}{2 \tau_i} \right) \nabla^{(1)} \cdot \tau_i \sum_\alpha \mathbf{e}_\alpha^i \mathbf{e}_\alpha^i S_\alpha^{i,(1)} \\ - \left(1 - \frac{\delta t}{2 \tau_i} \right) \nabla^{(1)} \cdot \tau_i \left[ \left( \partial_t^{(1)} \rho_i \mathbf{u} \right) \left( \mathbf{u} - \mathbf{u}^{eq}_i \right) + \left( \mathbf{u} - \mathbf{u}^{eq}_i \right) \left( \partial_t^{(1)} \rho_i \mathbf{u} \  \right) \right] \\ + \frac{\delta t}{2 \varepsilon} \partial_t^{(1)} \mathbf{\Theta}_i - \textbf{C}_{\textbf{p},i}^{(2)} + \textbf{C}_{\mathbf{\rho},i}^{(2)} + \frac{1}{\varepsilon} \textbf{C}_{\textbf{diff},i}^{(1)} + \mathcal{O}(Ma^3) \; ,
\end{aligned}
\end{equation}
where
\begin{equation} \label{eq:CE_theta_definition}
    \mathbf{\Theta}_i \equiv \sum_{j \neq i}^N \frac{\rho_i \rho_j}{\rho \tau_{ij}} \left(\mathbf{u}_i^{eq} - \mathbf{u}_j^{eq} \right)\; ,
\end{equation}
and the pressure, density and diffusion-driven contributions are
\begin{equation} \label{eq:CE_pressure_contribution}
\begin{aligned}
     \textbf{C}_{\textbf{p},i}^{(2)} = \left( 1 - \frac{\delta t}{2 \tau_i} \right) \nabla^{(1)} \cdot \tau_i \left[ \nabla^{(1)} \left(\rho_i c_{s,i}^2 \right) \mathbf{u} + \mathbf{u} \nabla^{(1)} \left( \rho_i c_{s,i}^2 \right) \right] \; ,
\end{aligned}
\end{equation}
\begin{equation} \label{eq:CE_density_contribution}
\begin{aligned}
     \textbf{C}_{\mathbf{\rho},i}^{(2)} = \left(1 - \frac{\delta t}{2 \tau_i} \right) c_{s,i}^2 \nabla^{(1)} \cdot \left[\tau_i \left( \nabla^{(1)} \rho \right) \cdot w_i \left( \mathbf{I} \mathbf{u}_i^{eq} +  \mathbf{u}_i^{eq} \mathbf{I} \right) \right]  \; ,
\end{aligned}
\end{equation}
\begin{equation} \label{eq:CE_diff_driven_contribution}
\begin{aligned}
     \textbf{C}_{\textbf{diff},i}^{(1)} = \left( 1 - \frac{\delta t}{2 \tau_i} \right) \nabla^{(1)} \cdot \left[ \frac{\tau_i \delta t}{2 \tau_i - \delta t} \left( \mathbf{\Theta}_i \mathbf{u} + \mathbf{u} \mathbf{\Theta}_i  \right) \right] \; .
\end{aligned}
\end{equation}

Hence, Eqs. (\ref{eq:CE_EQ59})--(\ref{eq:CE_EQ75}) represent the main derivations thus far and will be further explored in this section. Although the intermediate steps have not been shown here, we note that the second-order tensor $\mathbf{\Pi}^{(1)} \equiv \sum_\alpha \mathbf{e}_\alpha^i \mathbf{e}_\alpha^i f_\alpha^{i,(1)}$ is required in the algebraic manipulations needed to obtain Eq. (\ref{eq:CE_EQ75}). This tensor can be determined by evaluating the second-order moment of Eq. (\ref{eq:CE_EQ24}) and making the appropriate substitutions. The reader can find the algebraic details in the supplemental material. 

Multiplying Eqs. (\ref{eq:CE_EQ59}) and (\ref{eq:CE_EQ70}) by $\varepsilon$ and $\varepsilon^2$, combining them, and subsequently performing the reverse C-E expansion, the recovered mass conservation equation for the species $i$ reads
\begin{equation} \label{eq:CE_EQ78}
    \partial_t \rho_i + \nabla \cdot \rho_i \mathbf{u} = - \nabla \cdot \mathbf{j}_i\; ,
\end{equation}
where the mass diffusive flux $\mathbf{j}_i$ and the species velocity $\mathbf{u}_i$ must be
\begin{equation} 
    \mathbf{j}_i = \rho_i \left(\mathbf{u}_i^{eq} - \mathbf{u} \right) - \frac{\delta t}{2}  \mathbf{\Theta}_i  \; ,
\end{equation}
\begin{equation} \label{eq:CE_EQ83}
    \mathbf{u}_i = \mathbf{u}_i^{eq} - \frac{\delta t}{2 \rho_i} \mathbf{\Theta}_i  \; .
\end{equation}

Summing Eq. (\ref{eq:CE_EQ78}) upon all species $i$ and considering that $\tau_{ij} = \tau_{ji}$, $\rho = \sum_i \rho_i$, and $\mathbf{u} = \sum_i w_i \mathbf{u}_i^{eq}$, the recovered mass conservation equation for the mixture is
\begin{equation} \label{eq:CE_EQ84}
    \partial_t \rho + \nabla \cdot \rho \mathbf{u} = 0 \; .
\end{equation}

Similarly, by multiplying Eqs. (\ref{eq:CE_EQ67}) and (\ref{eq:CE_EQ75}) by $\varepsilon$ and $\varepsilon^2$, and combining them, we obtain
\begin{equation} \label{eq:CE_EQ89}
\begin{aligned}
    \left(\varepsilon \partial_t^{(1)} +  \varepsilon^2 \partial_t^{(2)} \right) \rho_i \mathbf{u}_i^{eq} + \varepsilon \nabla^{(1)} \cdot \left[\rho_i \left(\mathbf{u}_i^{eq} \mathbf{u} + \mathbf{u} \mathbf{u}_i^{eq} - \mathbf{u} \mathbf{u} \right) \right] = \\ - \varepsilon \nabla^{(1)} \rho_i c_{s,i}^2 + \left(1 - \frac{\delta t}{2 \tau_i} \right) c_{s,i}^2 \varepsilon \nabla^{(1)} \cdot \tau_i \rho \left[ \varepsilon \nabla^{(1)} w_i \mathbf{u}_i^{eq} + \left( \varepsilon \nabla^{(1)} w_i \mathbf{u}_i^{eq} \right)^\text{T} \right] \\ + \left(1 - \frac{\delta t}{2 \tau_i} \right) \varepsilon \nabla^{(1)} \cdot \left[ \tau_i \left( \mathbf{F}_i \mathbf{u} + \mathbf{u} \mathbf{F}_i \right) \right] - \left(1 - \frac{\delta t}{2 \tau_i} \right) \varepsilon \nabla^{(1)} \cdot \tau_i \sum_\alpha \mathbf{e}_\alpha^i \mathbf{e}_\alpha^i \varepsilon S_\alpha^{i,(1)} \\ - \left(1 - \frac{\delta t}{2 \tau_i} \right) \varepsilon \nabla^{(1)} \cdot \tau_i \left[ \varepsilon \left( \partial_t^{(1)} \rho_i \mathbf{u} \right) \left( \mathbf{u} - \mathbf{u}^{eq}_i \right) + \left( \mathbf{u} - \mathbf{u}^{eq}_i \right) \varepsilon \left( \partial_t^{(1)} \rho_i \mathbf{u} \right) \right] \\ + \frac{\delta t}{2} \varepsilon \partial_t^{(1)} \mathbf{\Theta}_i + \mathbf{F}_i - \mathbf{\Theta}_i + \varepsilon^2 \left( \textbf{C}_{\mathbf{\rho},i}^{(2)} - \textbf{C}_{\textbf{p},i}^{(2)} \right) + \varepsilon \textbf{C}_{\textbf{diff},i}^{(1)} + \mathcal{O}(Ma^3) \; .
\end{aligned}
\end{equation}
The time derivative is asymptotically expanded into a fast time scale $\partial_t^{(1)}$, characteristic of advective phenomena, and a slow time scale $\partial_t^{(2)}$, characteristic of diffusive phenomena, as shown in Eq. (\ref{eq:CE_time_derivative}). 
We assume that both $\left( \partial_t^{(1)} \rho_i \mathbf{u} \right) \left( \mathbf{u} - \mathbf{u}^{eq}_i \right)$ and $\partial_t^{(1)} \mathbf{\Theta}_i$ exhibit negligible fluctuations over the fast timescale and suggest that these two terms are unimportant in Eq. (\ref{eq:CE_EQ89}). This is reasonable, as these terms are associated with diffusive contributions that yield negligible variations on the fast time scale. Hence, the reverse C-E expansion yields the following conservation equation for species momentum, derived from the mesoscale viewpoint,
\begin{equation} \label{eq:epecies_momentum_eq}
\begin{aligned}
    \partial_t \rho_i \mathbf{u}_i^{eq} + \nabla \cdot \left[\rho_i \left(\mathbf{u}_i^{eq} \mathbf{u} + \mathbf{u} \mathbf{u}_i^{eq} - \mathbf{u} \mathbf{u} \right) \right] = - \nabla \rho_i c_{s,i}^2 \\ + \nabla \cdot \rho \nu_i \left[ \nabla w_i \mathbf{u}_i^{eq} + \left(  \nabla w_i \mathbf{u}_i^{eq} \right)^\text{T} \right] + \left(1 - \frac{\delta t}{2 \tau_i} \right) \nabla \cdot \left[ \tau_i \left( \mathbf{F}_i \mathbf{u} + \mathbf{u} \mathbf{F}_i \right) \right] \\ - \left(1 - \frac{\delta t}{2 \tau_i} \right) \nabla \cdot \tau_i \sum_\alpha \mathbf{e}_\alpha^i \mathbf{e}_\alpha^i S_\alpha^{i} \\ + \mathbf{F}_i - \mathbf{\Theta}_i + \textbf{C}_{\mathbf{\rho},i} - \textbf{C}_{\textbf{p},i} +  \textbf{C}_{\textbf{diff},i} + \mathcal{O}(Ma^3) \; ,
\end{aligned}
\end{equation}
where $\nu_i = \left(\tau_i - \delta t / 2 \right) c_{s,i}^2$.

Summing Eq. (\ref{eq:epecies_momentum_eq}) upon all the species $i$, some terms simplify or vanish. For instance, $\sum_i \mathbf{\Theta}_i = \mathbf{0}$, and $\sum_i \left( \textbf{C}_{\mathbf{\rho},i} - \textbf{C}_{\textbf{p},i} \right) = \mathbf{0}$. The term involving $\left( \mathbf{F}_i \mathbf{u} + \mathbf{u} \mathbf{F}_i \right)$ is a well-known spurious term that arises when recovering the NSE from force models in LBM and must be eliminated \cite{Luo_forcing_scheme, Luo_2000}. As the term containing $S_\alpha^{i}$ acts as a corrective factor to remove any spurious terms, the standard forcing approach, derived from a second-order expansion of the forcing term in velocity space, is capable of removing it \cite{Luo_forcing_scheme},
\begin{equation} \label{eq:CE_forcing_term_discretized}
    S_\alpha^i = \omega_\alpha \left[\frac{\mathbf{e}_\alpha^i}{c_{s,i}^2} + \frac{\left(\mathbf{e}_\alpha^i \mathbf{e}_\alpha^i - c_{s,i}^2 \mathbf{I} \right)}{c_{s,i}^4} \cdot \mathbf{u} \right] \cdot \mathbf{F}_i \; .
\end{equation}
Alongside these considerations, by assuming the ideal equation of state $p_i = \rho_i c_{s,i}^2$, the total pressure as the sum of the partial pressures, i.e., $p=\sum_i p_i$, and $\mathbf{F}=\sum_i \mathbf{F}_i$, the NSE is recovered, 
\begin{equation} \label{eq:NSE}
\begin{aligned}
    \partial_t \rho \mathbf{u} + \nabla \cdot \rho \mathbf{u} \mathbf{u} = - \nabla p + \nabla \cdot \rho \nu \left[ \nabla  \mathbf{u} + \left( \nabla \mathbf{u} \right)^\text{T} \right] + \mathbf{F}  \; .
\end{aligned}
\end{equation}

In contrast to the work of Tong et al. \cite{Tong_etal2014}, the assumption $-\sum_i \rho_i \mathbf{u}_i^{eq} \mathbf{u}_i^{eq} \approx -\rho \mathbf{u} \mathbf{u}$ is unnecessary here, as $\rho_i \left(\mathbf{u}_i^{eq} \mathbf{u} + \mathbf{u} \mathbf{u}_i^{eq} - \mathbf{u} \mathbf{u} \right)$ in Eq. (\ref{eq:epecies_momentum_eq}) results in  $-\rho \mathbf{u} \mathbf{u}$ when the summation is employed. Additionally, the viscous contribution in their work was recovered only under the inconsistent supposition that $\sum_i \rho_i \nu_i \left[ \nabla \mathbf{u}_i^{eq} + \left( \nabla \mathbf{u}_i^{eq} \right)^\text{T} \right] \approx \rho \nu \left[ \nabla \mathbf{u} + \left( \nabla \mathbf{u} \right)^\text{T} \right]$, where $\nu_i$ was physically interpreted as the contribution of each species to the mixture kinematic viscosity. However, this hypothesis is also unnecessary here because the constraint $\mathbf{u} = \sum_i w_i \mathbf{u}_i^{eq}$ can be enforced directly, though under the assumption that $\nu_i = \nu \; \forall \; i$. Although the previous physical interpretation of $\nu_i$ is now lost, this approach allows for incorporating viscosity mixing rules, such as $\nu = f(x_i)$, where $x_i$ is the mole fraction.

Unlike the mixture momentum equation (namely the NSE), the macroscopic viewpoint does not provide a separate momentum equation for each species. Consequently, it is not possible to directly compare Eq. (\ref{eq:epecies_momentum_eq}) with a corresponding equation at the macroscale to identify discrepancies or enhance the physical interpretation of each term. Given that the M-S closure equation for mass transfer modeling originates from a momentum balance \cite{Taylor&Krishna}, it may be related to the previously derived momentum equations. With this in mind, not only will the term containing $\left( \mathbf{F}_i \mathbf{u} + \mathbf{u} \mathbf{F}_i \right)$ act as a spurious term in Eq. (\ref{eq:epecies_momentum_eq}), but the contributions $\textbf{C}_{\textbf{p},i}$, $\textbf{C}_{\mathbf{\rho},i}$, and $\textbf{C}_{\textbf{diff},i}$ will as well. Hence, rigorous modeling indicates that the forcing term should be refined to eliminate them, which gives rise to
\begin{equation} \label{eq:improved_forcing_scheme}
    S_\alpha^i = \omega_\alpha \left[\frac{\mathbf{e}_\alpha^i}{c_{s,i}^2} + \frac{\left(\mathbf{e}_\alpha^i \mathbf{e}_\alpha^i - c_{s,i}^2 \mathbf{I} \right)}{c_{s,i}^4} \cdot \mathbf{u} \right] \cdot \mathbf{F}_i - S_{\text{p},\alpha}^i + S_{\rho,\alpha}^i + S_{\text{diff},\alpha}^i\; ,
\end{equation}
where the pressure, density and diffusion-driven forcing contributions are
\begin{equation}
    S_{\text{p},\alpha}^i = \omega_\alpha  \left[ \frac{\left(\mathbf{e}_\alpha^i \mathbf{e}_\alpha^i - c_{s,i}^2 \mathbf{I} \right)}{c_{s,i}^4} \cdot \mathbf{u} \right] \cdot  \nabla \rho_i c_{s,i}^2  \; ,
\end{equation}
\begin{equation}
    S_{\rho,\alpha}^i = \omega_\alpha \left[ \frac{\left(\mathbf{e}_\alpha^i \mathbf{e}_\alpha^i - c_{s,i}^2 \mathbf{I} \right)}{c_{s,i}^2} \cdot \mathbf{u}^{eq}_i \right] \cdot \left( w_i \nabla \rho \right)  \; ,
\end{equation}
\begin{equation}
    S_{\text{diff},\alpha}^i = \omega_\alpha \left(\frac{\delta t}{2 \tau_i - \delta t}\right) \left[ \frac{\left(\mathbf{e}_\alpha^i \mathbf{e}_\alpha^i - c_{s,i}^2 \mathbf{I} \right)}{c_{s,i}^4} \cdot \mathbf{u} \right] \cdot \mathbf{\Theta}_i  \; .
\end{equation}

Note that the forcing contributions discussed above will only affect the recovered species momentum equation and any subsequent equations derived from it, leaving the recovered NSE unchanged however. Multiplying Eqs. (\ref{eq:epecies_momentum_eq}) and (\ref{eq:NSE}) by $1/p$ and $w_i/p$, respectively, and then subtracting Eq. (\ref{eq:NSE}) from Eq. (\ref{eq:epecies_momentum_eq}),
\begin{equation} \label{eq:CE_EQ94}
\begin{aligned}
    \frac{1}{p} \left( \partial_t \rho_i \mathbf{u}_i^{eq} - w_i \partial_t \rho \mathbf{u} \right) + \frac{1}{p} \left[ \nabla \cdot \rho_i \left( \mathbf{u}_i^{eq} \mathbf{u} + \mathbf{u} \mathbf{u}_i^{eq} - \mathbf{u} \mathbf{u} \right) - w_i \nabla \cdot \rho \mathbf{u} \mathbf{u} \right] \\ - \frac{1}{p} \left[  \nabla \cdot \rho \nu_i \left( \nabla w_i \mathbf{u}_i^{eq} + \left( \nabla w_i \mathbf{u}_i^{eq} \right)^\text{T} \right) - w_i \nabla \cdot \rho \nu \left( \nabla \mathbf{u} + \left( \nabla \mathbf{u} \right)^\text{T} \right) \right] = \\ - \frac{1}{p} \nabla p_i + \frac{w_i}{p} \nabla p  - \frac{1}{p} \mathbf{\Theta}_i + \frac{1}{p} \left(\mathbf{F}_i - w_i \mathbf{F} \right)  \; .
\end{aligned}
\end{equation}
As established in the development of the two-fluid theory from the kinetic perspective, all terms on the left-hand side of Eq. (\ref{eq:CE_EQ94}) can be neglected by assuming that those derivatives vary slowly on the maxwellization  ($f^i_\alpha \rightarrow f^{i(0)}_\alpha$) time scale \cite{Goldman&Sirovich}. Then, recalling the definition of $\mathbf{\Theta}_i$ from Eq. (\ref{eq:CE_theta_definition}), considering that $p_i = p x_i$, rearranging the terms, and defining $\mathbf{F}_i = \rho_i \mathbf{k}_i$, the M-S equation is recovered for ideal mixtures,
\begin{equation} \label{eq:MS_equation}
     - \sum_{j \neq i}^N \frac{\rho_i \rho_j}{\rho p \tau_{ij}} \left(\mathbf{u}_i^{eq} - \mathbf{u}_j^{eq} \right) = \nabla x_i + \left( x_i - w_i \right) \frac{\nabla p}{p} - \frac{\rho_i}{p} \left(\mathbf{k}_i - \sum_{j=1}^N w_j \mathbf{k}_j \right)  \; ,
\end{equation}
where $\mathbf{k}_i$ represents the specific force acting on species $i$.

Eq. (\ref{eq:MS_equation}) is compared to the well-known M-S equation for ideal mixtures to determine the relationship between $\tau_{ij}$ and the diffusion coefficient ${\text{\DJ}}_{ij}$ of the pair $i$-$j$, which imposes that 
\begin{equation} \label{eq:CE_EQ102}
     \frac{\rho_i \rho_j}{\rho p \tau_{ij}} \left(\mathbf{u}_i^{eq} - \mathbf{u}_j^{eq} \right) = \frac{x_i x_j}{\text{\DJ}_{ij}} \left(\mathbf{u}_i - \mathbf{u}_j \right)   \; .
\end{equation}
Substituting the real species velocity $\mathbf{u}_i$ as presented in Eq. (\ref{eq:CE_EQ83}), along with $x_i = \rho_i/n M_i$, where $n$ is the total mole number, and performing some algebraic manipulations, the connection between $\tau_{ij}$ and ${\text{\DJ}}_{ij}$ comes forth,
\begin{equation} \label{eq:CE_EQ107}
    \begin{aligned}
   \left( \frac{\rho_i + \rho_j}{\text{\DJ}_{ij}} + \frac{2 n^2 M_i M_j}{p \delta t}\right) \frac{1}{\tau_{ij}} + \frac{1}{\text{\DJ}_{ij}} \sum^N_{s \neq i,j} \left[ \frac{\rho_s}{\tau_{si}} + \left( \frac{\rho_s}{\tau_{sj}} - \frac{\rho_s}{\tau_{si}} \right) \left( \frac{\mathbf{u}_j^{eq} - \mathbf{u}_s^{eq}}{\mathbf{u}_j^{eq} - \mathbf{u}_i^{eq} } \right) \right] \\ = \frac{2\rho}{\text{\DJ}_{ij} \delta t} \; ,
   \end{aligned}
\end{equation}
which differs from the relationship obtained in the work of Tong et al. \cite{Tong_etal2014}. The most significant distinction is the necessity of including the species velocity field in the calculation of $\tau_{ij}$. Nevertheless, Eq. (\ref{eq:CE_EQ107}) remains consistent with the approach by Luo and Girimaji \cite{LuoGirimaji2002,LuoGirimaji2003}, as it accurately reduces to the established relationship for binary mixtures,
\begin{equation} \label{eq:Luo&Girimaji_tau}
   \tau_{ij} = \frac{\delta t}{2} + \frac{n^2 M_i M_j}{p \rho} \text{\DJ}_{ij}  \; .
\end{equation}

However, maintaining the velocity field dependence in Eq. (\ref{eq:CE_EQ107}) is unfavorable as it may compromise numerical stability, particularly near the system equilibration ($\mathbf{u}^{eq}_i \rightarrow \mathbf{u}^{eq}_j$), where differences in species velocities disappear. Hence, under the assumption that the species velocities differ by the same order, Eq. (\ref{eq:CE_EQ107}) simplifies to
\begin{equation} \label{eq:relation_tauij_diff_coeff}
    \begin{aligned}
   \left( \frac{\rho_i + \rho_j}{\text{\DJ}_{ij}} + \frac{2 n^2 M_i M_j}{p \delta t}\right) \frac{1}{\tau_{ij}} + \frac{1}{\text{\DJ}_{ij}} \sum^N_{s \neq i,j} \frac{\rho_s}{\tau_{sj}}  = \frac{2\rho}{\text{\DJ}_{ij} \delta t} \; ,
   \end{aligned}
\end{equation}
which can be iteratively solved in a multicomponent system to obtain a set of $\tau_{ij}$ that aligns with the corresponding set of $\text{\DJ}_{ij}$.

To conclude our discussion on the C-E expansion, this section analyzed the LBE using the proposed forcing approach within the EVD model. Specifically, we considered Eq.  (\ref{eq:LBE_forcing}) instead of Eq. (\ref{eq:LBE_Sirovich}), with the proposed forcing term defined by Eq. (\ref{eq:improved_forcing_scheme}). In contrast to the methodology described in Section \ref{sec:theory}, the equilibrium velocity must be determined by Eq. (\ref{eq:first_order_moment_force}) to ensure second-order space–time accuracy when external forces are taken into account.

\section{Numerical simulations}\label{sec:simulations}

In this section, we investigate the proposed forcing approach for simulating mass transfer under the influence of external forces. The analysis is divided into three sections, each addressing a specific aspect of the validation and applicability of the proposed approach. We first show that it effectively accounts for relevant forcing effects by simulating the dynamics of a ternary miscible mixture under the influence of a gravitational force field in Section \ref{sec:gravitational_force}, a type of problem ubiquitous in real-world scenarios, such as compositional grading in oil reservoirs. In Section \ref{sec:centrifugal_force}, we demonstrate that the proposed approach not only handles forcing effects but also accurately recovers the expected analytical solutions for typical mass transport scenarios. This is achieved by examining the ultracentrifugation of uranium isotopes, which offers a known analytical solution to serve as a robust basis for comparison. Lastly, Section \ref{sec:Couette} examines the role of the corrective forcing contributions outlined in Eq. (\ref{eq:improved_forcing_scheme}), specifically $S_{\text{p},\alpha}^i$, $S_{\rho,\alpha}^i$, and $S_{\text{diff},\alpha}^i$. The permeable Couette flow benchmark is implemented. The analysis indicates that, while these corrective contributions may be necessary for a rigorous recovery of the macroscopic equations, their impact is negligible in certain cases.

In all simulations, the D2Q9 lattice arrangement is employed with $\delta x = 1$ and $\delta t = 1$, incorporating the DLS scheme \cite{McCracken2005} and second-order spatial interpolations \cite{Tong_etal2014}. We set $M_1 = 1$ and $c_1 = 1$ for the lightest species, $\rho_{i_0} = 1$ for the heaviest species, and $\nu_i = \nu_1 \; \forall \; i$. The remaining implementation details are provided in each section.

\subsection{Case I: Loschmidt tube with gravitational force}\label{sec:gravitational_force}

\begin{figure}[h]
    	\centering
    	\includegraphics[width=0.55\linewidth]{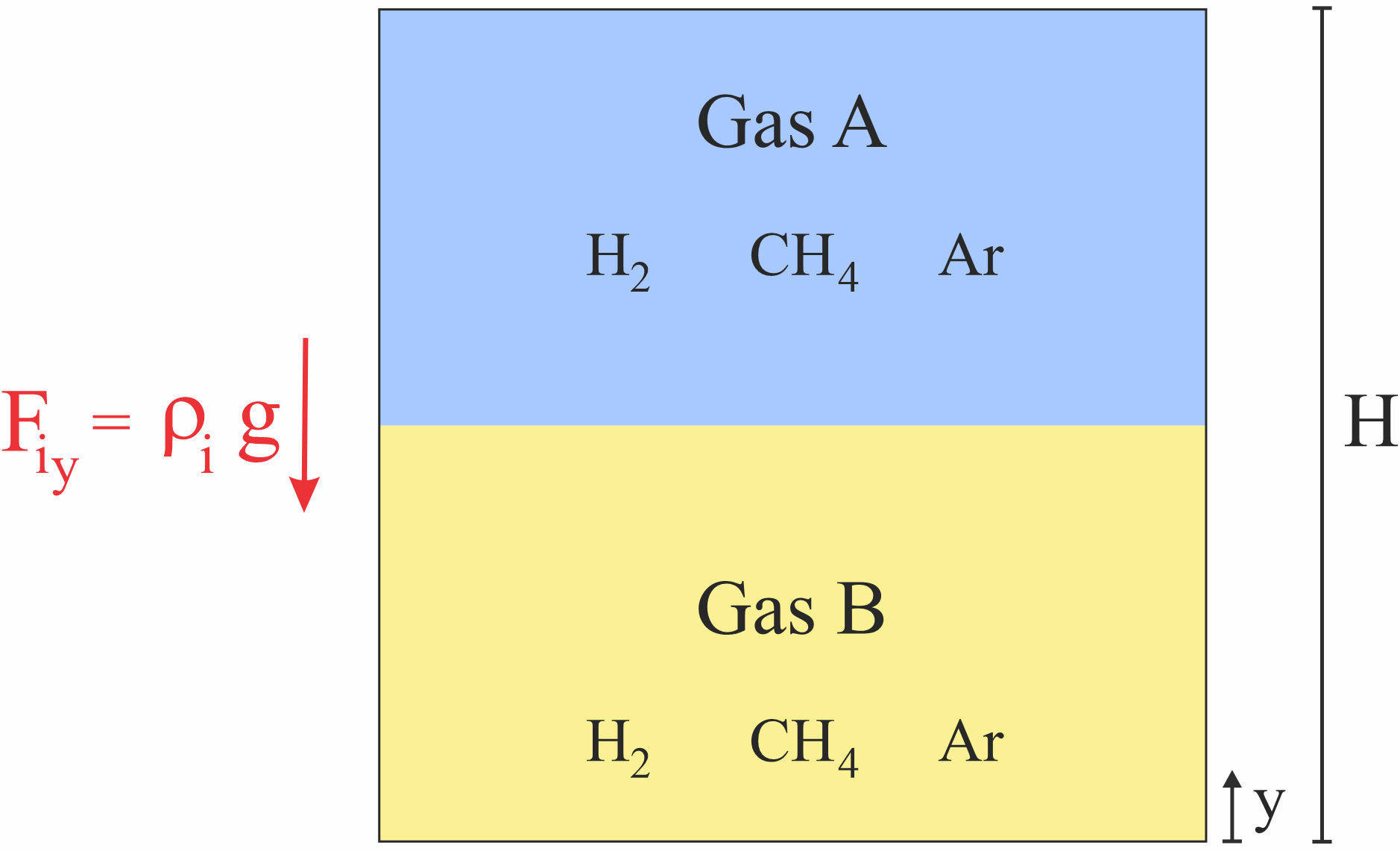}
    	\caption{Sketch of the two-dimensional ternary mixture simulated in the Loschmidt tube, where H denotes the domain height.}
    	\label{fig:Sketch_Loschmidt}
    \end{figure}

We implemented the Loschmidt tube benchmark to initially assess the applicability of the proposed forcing approach. It is a widely used test for diffusion modeling in miscible mixtures, but here, we introduce a gravitational force field, which is not commonly part of this benchmark. The ternary mixture composed of H$_2$ (1), CH$_4$ (2), and Ar (3) is set at 307.15 K and 101.3 kPa. As illustrated in Fig. \ref{fig:Sketch_Loschmidt}, the tube is a square grid (200 $\times$ 200 nodes) with a height of $H = 1$ m. A gravitational force $F_{i_y} = \rho_i g$ acts on each species in the y-direction, with the specific force $g$ varying across the simulated cases as $g = [0, 10^{-7}, 10^{-6}, 10^{-5}]$ l.u. This corresponds to Galilei numbers ($Ga = g H^3/\nu^2$) in the range $Ga = [0, 30, 300, 3000]$. The gases A ($x_1 = 0.001$ and $x_3 = 0.7$) and B ($x_1 = 0.8$ and $x_3 = 0.001$) are initialized with zero velocity. The molecular masses of the pure components and the diffusion coefficients, known from the literature and presented in Tables \ref{table:Data_Validation_Sirovich} and \ref{table:DiffusionCoeffDataLoschmidt}, are incorporated into the implementations. The densities of the pure components $\rho_{i_0}$ are calculated using the Virial equation of state. After conveniently setting $\tau_1 = 1$, the remaining LBM variables are determined through similarity principles. The total mole number is initialized based on the local composition of each species using $n = \sum_i x_i \rho_{i_0}/M_i$, from which the initial species and fluid densities emerge, $\rho_i = x_i M_i n$ and $\rho = \sum_i \rho_i$. The cross-relaxation parameters are tuned by analytically solving the linear algebraic system derived from Eq. (\ref{eq:relation_tauij_diff_coeff}) for a ternary setup at each time step and domain position,
\begin{equation} 
        \tau_{23} =  \left(A_{23} - \frac{\rho_1 \rho_2}{A_{13} \text{\DJ}_{13} \text{\DJ}_{23}} \right) \left(B_{23} - \frac{\rho_1 B_{13}}{A_{13} \text{\DJ}_{23} } \right)^{-1} \; ,
\end{equation}
\begin{equation} 
        \tau_{12} =  \left( \frac{B_{12}}{A_{12}} -  \frac{\rho_3}{A_{12} \text{\DJ}_{12} \tau_{23}}  \right)^{-1} \; ,
\end{equation}
\begin{equation} 
        \tau_{13} =  \left( \frac{B_{13}}{A_{13}} -  \frac{\rho_2}{A_{13} \text{\DJ}_{13} \tau_{23}}  \right)^{-1} \; ,
\end{equation}
where
\begin{equation} 
        A_{ij} = \frac{\rho_i + \rho_j}{\text{\DJ}_{ij}} + \frac{2 n^2 M_i M_j}{p \delta t} \; ,
\end{equation}
$B_{ij} = 2 \rho / \left(\text{\DJ}_{ij} \delta t\right)$, and the pressure of an ideal mixture can be assigned as $p = n M_1 c_{s,1}^2$ within the EVD model.

\begin{table}[h]
    \centering
    \caption{Molecular mass \cite{SmithVanNess&Abbott2004} and density of pure components at 307.15 K and 101.3 kPa.}
    \label{table:Data_Validation_Sirovich}
    \begin{tabular}{c c c}
    \hline
    Species &
    $M_i$ (g/mol) &
    $\rho_{i_0}$ (kg/m$^3$) \\
    \hline
    H$_2$  & 2.016          & 0.080  \\
    CH$_4$ & 16.043         & 0.638  \\
    Ar   & 39.948         & 1.586    \\ 
    \hline
    \end{tabular}
\end{table}

\begin{table}[h]
    \centering
    \caption{Diffusion coefficients for the pairs composed of H$_2$ (1), CH$_4$ (2), and Ar (3) \cite{Taylor&Krishna}.}
    \label{table:DiffusionCoeffDataLoschmidt}
    \begin{tabular}{c c c}
    \hline
    $\text{\DJ}_{12}$ (m$^2$/s) & $\text{\DJ}_{13}$ (m$^2$/s) & $\text{\DJ}_{23}$ (m$^2$/s) \\ 
    \hline
    77.16 $\times$ 10$^{-6}$    & 83.35 $\times$ 10$^{-6}$    & 21.57 $\times$ 10$^{-6}$    \\ 
    \hline
    \end{tabular}
\end{table}

\begin{figure}[htbp]
    	\centering
    	\includegraphics[width=0.55\linewidth]{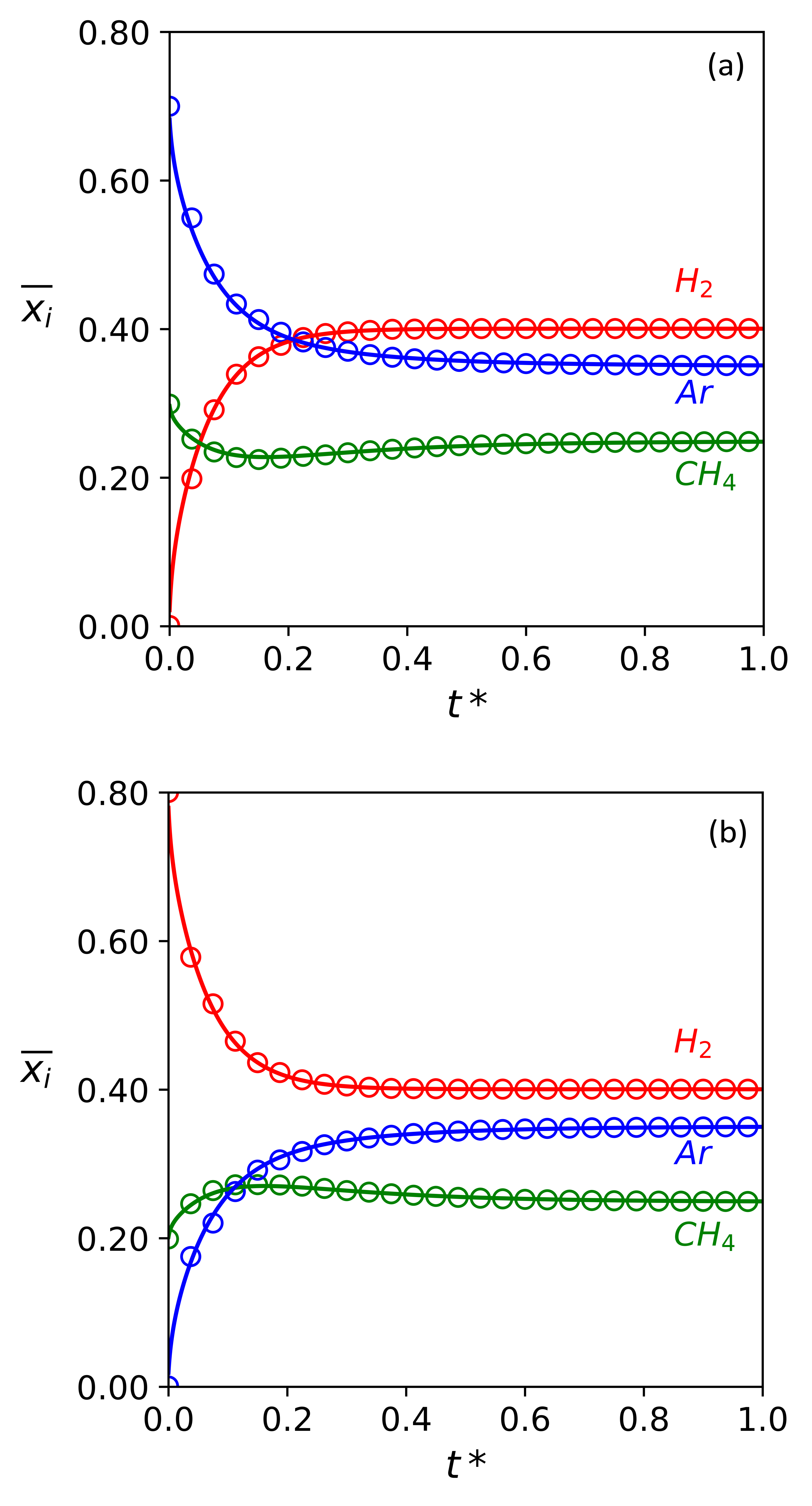}
    	\caption{Average mole fractions ($\overline{x_i}$) in the top (a) and bottom (b) regions of the Loschmidt tube are depicted over dimensionless time $t^*$ for $Ga = 0$. This simulation will guide the discussion for the cases where $Ga \neq 0$. Points represent the numerical solution using LBM and lines represent the solution through the linearized theory. The reference time for nondimensionalization is set to 40,000 steps.}
    	\label{fig:Loschmidt_analytical_comparison}
    \end{figure}

Since the Loschmidt tube with a force field lacks an available analytical solution, we will first establish the case of $Ga = 0$ as the reference simulation, which offers an analytical solution based on the standard linearized theory \cite{Taylor&Krishna}. Fig. \ref{fig:Loschmidt_analytical_comparison} shows that the LBM implementation closely matches the expected profiles of the average composition $\overline{x_i}$ of each species over time, where $\overline{x_i} = \sum_{\mathbf{x}} x_i(\mathbf{x}) / (0.5 \times 200^2)$ is evaluated separately for the top ($H/2 \leq y \leq H$) and bottom ($0 \leq y \leq H/2$) regions of the tube. This alignment indicates that the investigated methodology preserves both accuracy and physical consistency, and that Eq. (\ref{eq:relation_tauij_diff_coeff}), obtained in the C-E analysis, accurately adjusts $\tau_{ij}$ to correspond with the specified $\text{\DJ}_{ij}$.

\begin{figure}[h]
    	\centering
    	\includegraphics[width=0.7\linewidth]{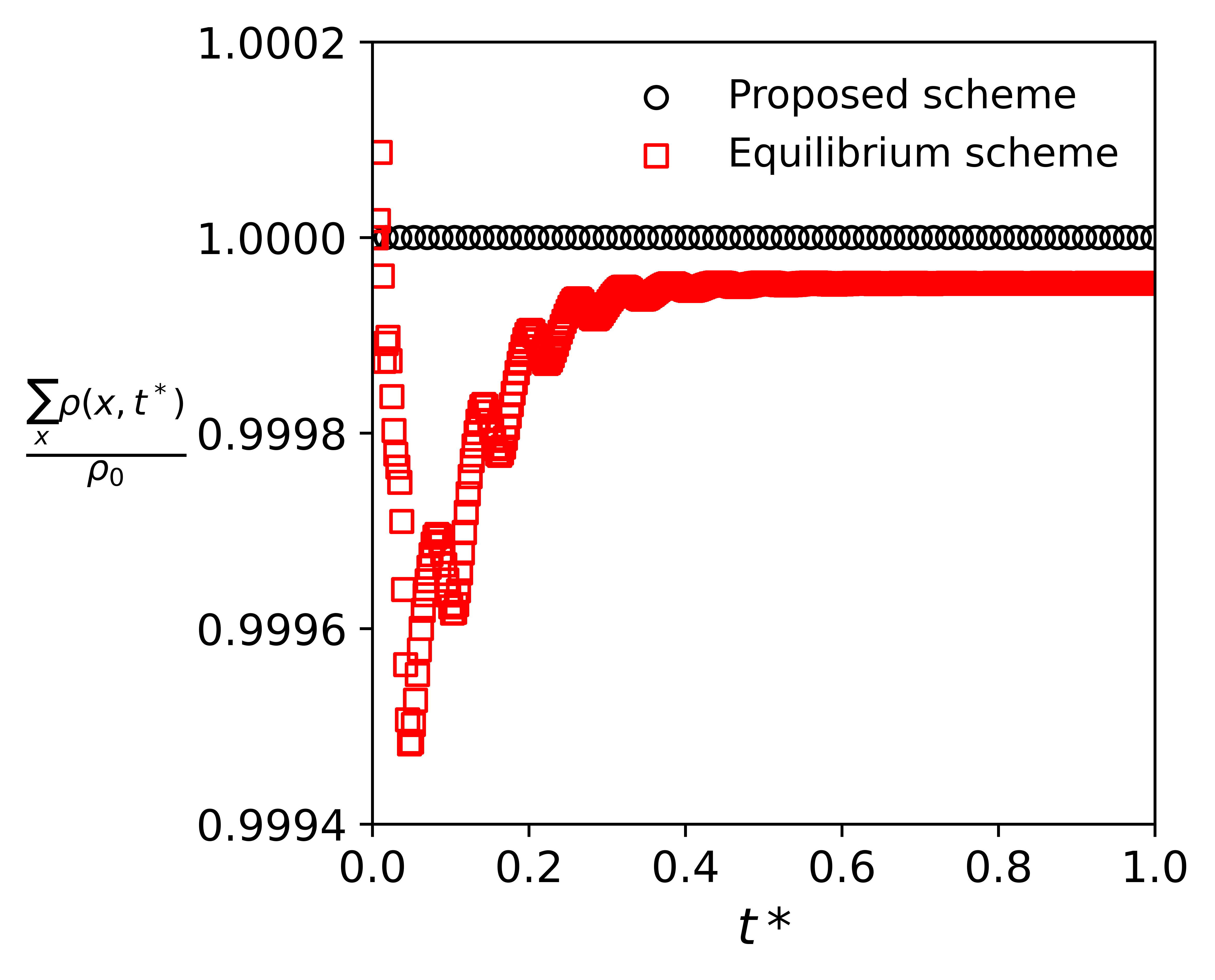}
    	\caption{Total mass in the Loschmidt tube, normalized by the initial total mass $\rho_0 = \sum_{\textbf{x}} \rho(\textbf{x},t^*=0)$, is plotted over dimensionless time $t^*$ for two considered boundary schemes. The reference time for nondimensionalization is set to 40,000 steps.}
    	\label{fig:Loschmidt_boundary_comparison}
    \end{figure}

Note that usual periodic conditions are applied to the lateral boundaries, while a novel boundary scheme proposed in \ref{sec:bouce_back} is implemented at the upper and lower boundaries. This new formulation is necessary to ensure mass conservation throughout the simulated domain, as the standard halfway bounce-back scheme fails to maintain mass conservation in mixtures with varying molecular masses, where spatial interpolations are required. Using the previous simulation with $Ga=0$ as a demonstration, Fig. \ref{fig:Loschmidt_boundary_comparison} illustrates the evolution of the total mass over time within the domain and confirms the proposed scheme ensures mass conservation while maintaining stationary, solid, and impermeable boundary characteristics. Using a single distribution function for both mass and momentum transport introduces challenges in boundary implementations. The equilibrium scheme, for instance, facilitates this process by enabling the quick specification of distributions based on the imposed macroscopic properties of the species and mixture. Here, the reference case ($Ga=0$) was also examined using the equilibrium scheme, where $f_\alpha^i = f_\alpha^{i,(0)} (\rho_{\text{s}},\mathbf{u}=0)$ in the impermeable boundaries, with $\rho_\text{s}$ representing the density of the adjacent fluid node. Fig. \ref{fig:Loschmidt_boundary_comparison} demonstrates that the equilibrium scheme fails to ensure mass conservation. In contrast, the proposed boundary scheme guarantees that the total mass remains constant from initialization, thereby ensuring mass conservation as theoretically predicted in \ref{sec:bouce_back}.
 
\begin{figure}[h]
    	\centering
    	\includegraphics[width=1.\linewidth]{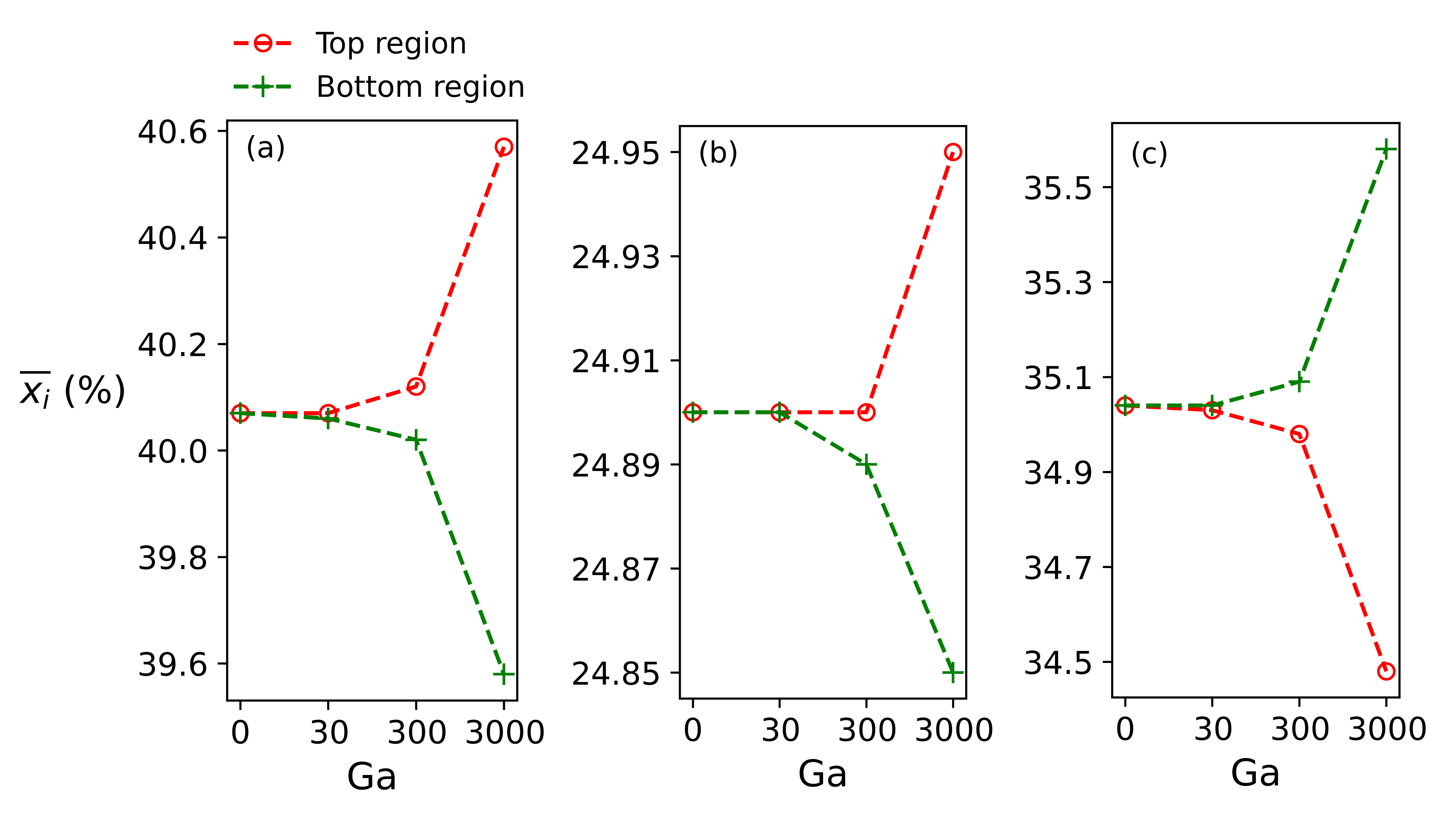}
    	\caption{Percent average mole concentration at steady-state for different values of $Ga$ for (a) H$_2$, (b) CH$_4$, and (c) Ar. The lines are plotted to help guide the eyes. Note that the vertical scales differ to account for variations in the simulated concentrations of each species.}    	\label{fig:Loschmidt_tube_force_concentration}
    \end{figure}

Having established $Ga = 0$ as the reference case, we now turn to cases where $Ga \neq 0$. The steady-state mole concentrations for each species are displayed for the considered $Ga$ range in Figs. \ref{fig:Loschmidt_tube_force_concentration}a to \ref{fig:Loschmidt_tube_force_concentration}c. For $Ga=0$, the concentrations are identical at the top and bottom regions for each species, representing the equilibrium condition achieved when no external force contributes to the M-S equation. As the gravitational force grows, the concentrations in the two regions diverge, reaching the maximum deviation at $Ga=3000$, the highest value investigated. For $Ga=0$, the H$_2$ and Ar concentrations naturally increase over time at the top and bottom regions, respectively, due to the pressure and concentration contributions to the diffusive driving force in the M-S equation. When $Ga \neq 0$, H$_2$ (the lightest species) becomes more concentrated in the top, while Ar (the heaviest species) in the bottom, due to the differences in molecular masses and the presence of the gravitational field, as presented in Figs. \ref{fig:Loschmidt_tube_force_concentration}a and \ref{fig:Loschmidt_tube_force_concentration}c. In contrast, CH$_4$ has an intermediate molecular mass, making its distribution within the domain harder to predict in advance. Fig. \ref{fig:Loschmidt_tube_force_concentration}b reveals that CH$_4$  tends to accumulate in the top region, similar to H$_2$, but specifically for $Ga > 30$.

\begin{figure}[h]
    	\centering
    	\includegraphics[width=1.\linewidth]{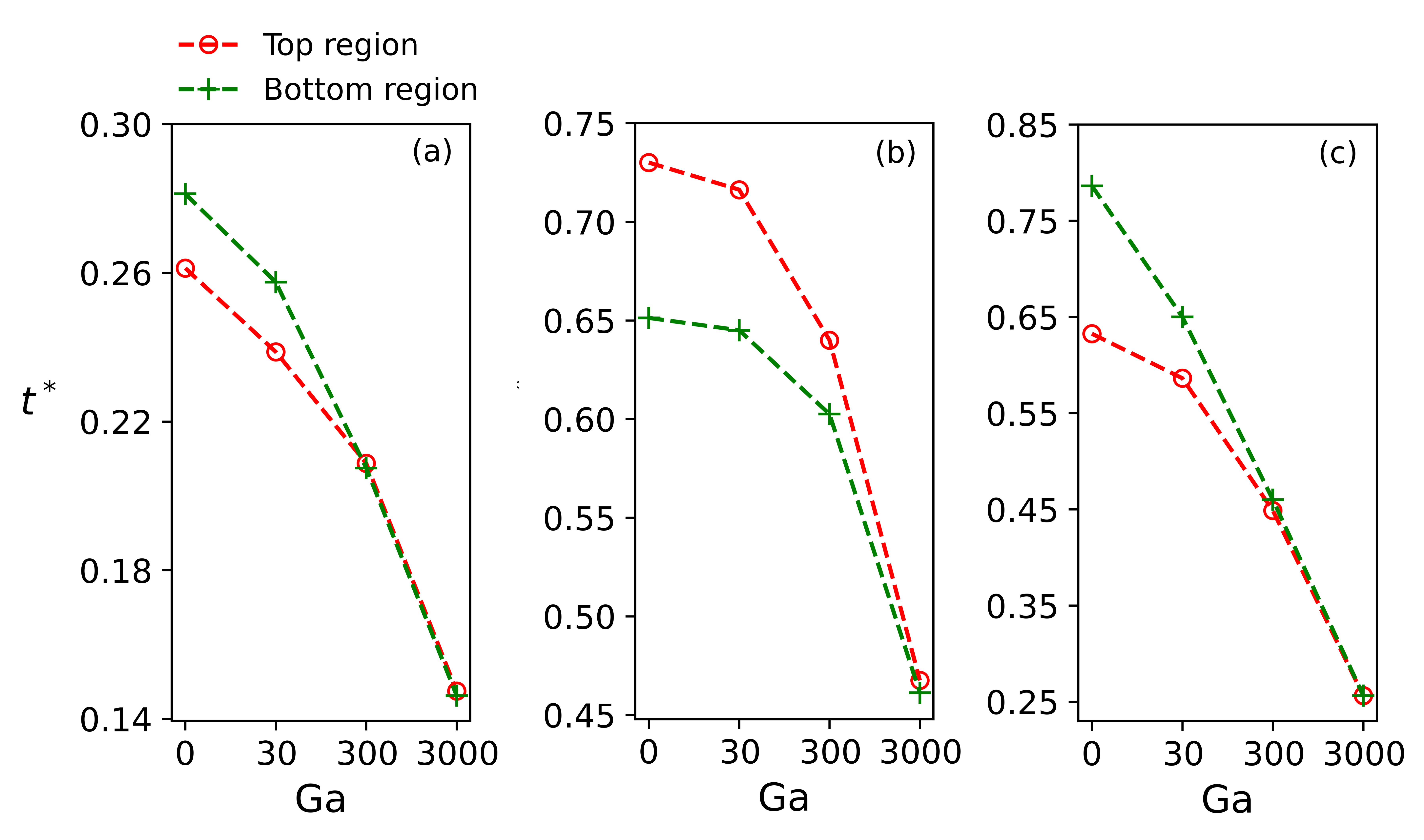}
    	\caption{Dimensionless time required to reach the same steady-state concentration as in the reference case ($Ga=0$) for different values of $Ga$. The results are shown for (a) H$_2$, (b) CH$_4$, and (c) Ar. The reference time for nondimensionalization is set to 80,000 steps. The lines are plotted to help guide the eyes.}    	  \label{fig:Loschmidt_tube_force_time}
    \end{figure}

The time $t^*$ required for the system to reach the steady-state concentrations varies by species and region within the tube, as observed in Fig. \ref{fig:Loschmidt_tube_force_time}. For instance, it takes $t^*\approx 0.26$ and $t^*\approx 0.28$ for H$_2$ to reach its equilibrium concentration for $Ga=0$ in the top and bottom regions, respectively, whereas it takes a longer time for Ar reach its equilibrium concentration under the same conditions ($t^*\approx 0.63$ and $t^*\approx 0.79$). Note that we set the reference time for nondimensionalization to 80,000 steps. The system dynamics change when $Ga>0$, leading to the same equilibrium concentrations observed in the reference simulation being reached more rapidly, with shorter $t^*$ values. This effect is particularly pronounced for $Ga=3000$, as the required $t^*$ for each tube region becomes very similar ($t^* \approx 0.15$ for H$_2$, $t^* \approx 0.46$ for CH$_4$, and $t^* \approx 0.26$ for Ar).

Obviously, the gravity effect should become more pronounced when either the gravity field is stronger or the density difference between components is larger. But as proof of concept, we chose this relatively simple system to show that the forcing scheme proposed here can address this effect, which previous EVD-LBM models were unable to, while still maintaining a multifluid perspective rather than the single-fluid approach employed in the passive scalar models. We emphasize that this type of problem is particularly relevant to simulations of compositional grading in oil reservoirs and geological storage of CO$_2$, where gravity directly impacts the distribution of components along reservoirs. The current benchmark already establishes the applicability of the proposed forcing approach; however, to provide a proper basis for evaluation, we implement the ultracentrifuge benchmark in Section \ref{sec:centrifugal_force} to compare the concentration profiles with the analytical results when external forces are considered.

\subsection{Case II: ultracentrifugation of uranium isotopes}\label{sec:centrifugal_force}

\begin{figure}[h]
    	\centering
    	\includegraphics[width=0.55\linewidth]{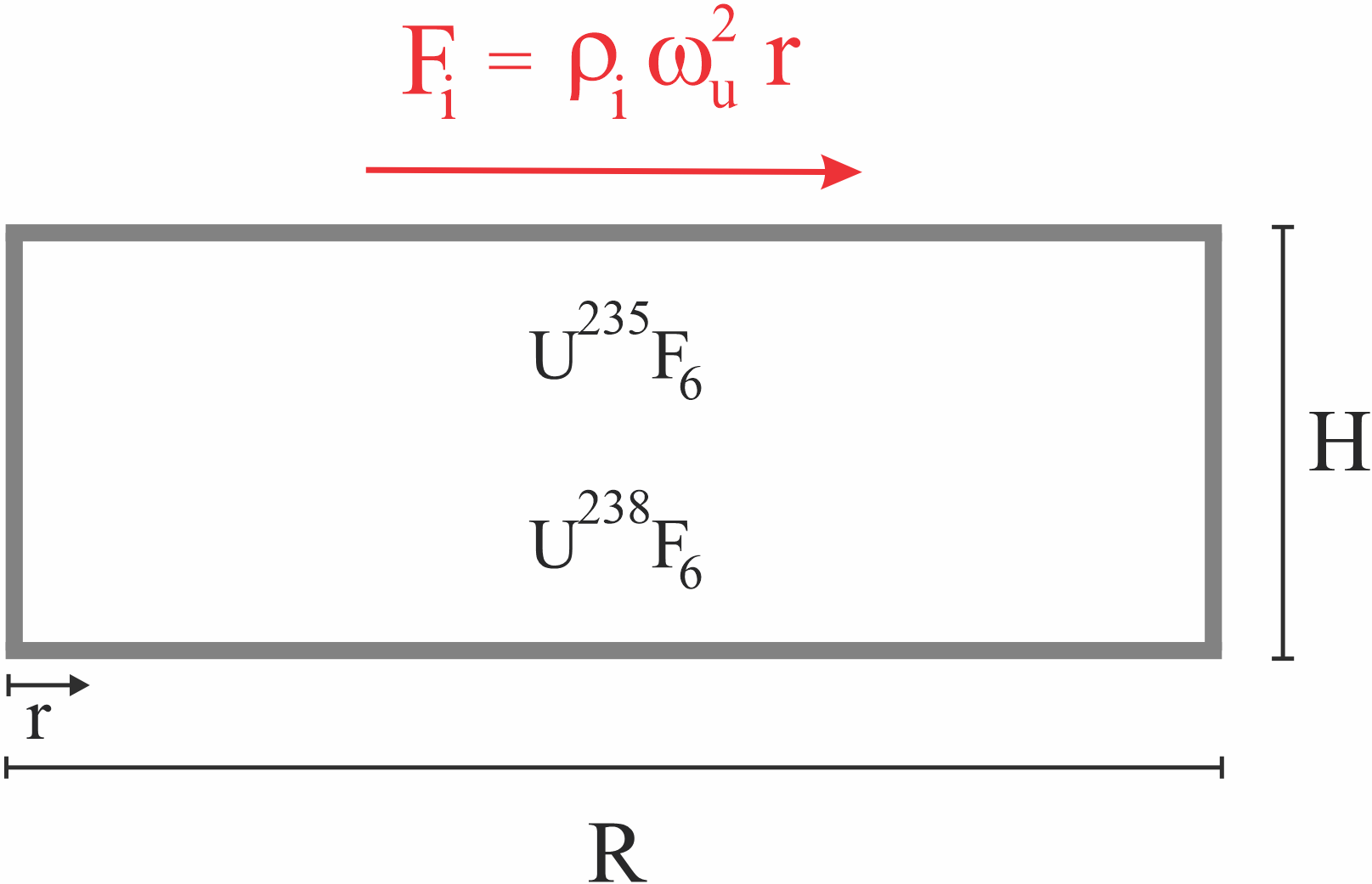}
    	\caption{Sketch of the binary diffusion in the two-dimensional ultracentrifuge, where R denotes the radius.}
    	\label{fig:Sketch_ultracentrifuge}
    \end{figure}

This benchmark is based on ultracentrifuge isotope separation and aims to validate the proposed scheme by comparing simulated concentration profiles with analytical results. It features a rectangular two-dimensional closed cavity ($H = 3$ nodes and $R = 500$ nodes) with stationary walls, subjected to a centrifugal force $F_{i}(r) = \rho_i(r) {\omega_u}^2 r$ in the $r$-direction, as illustrated in Fig. \ref{fig:Sketch_ultracentrifuge}. The analytical solution provides the steady-state separation factor $\alpha(r)$ and concentration profile $x_1(r)$ \cite{Taylor&Krishna},
\begin{equation} \label{eq:separation_factor_eq}
        \frac{1}{\alpha} \equiv \frac{x_1(1-x_{1}|_0)}{x_{1}|_0(1-x_1)} = \exp \left( \frac{\mathbb{C} r^2}{2}\right)  \; .
\end{equation}
Solving Eq. (\ref{eq:separation_factor_eq}) for $x_1$ results in
\begin{equation} 
        x_1(r) = \frac{\left(\frac{x_{1}|_0}{1-x_{1}|_0} \right) \exp\left(\frac{\mathbb{C} r^2}{2} \right)}{1 + \left(\frac{x_{1}|_0}{1-x_{1}|_0} \right) \exp\left(\frac{\mathbb{C} r^2}{2} \right)}  \; ,
\end{equation}
where $x_1|_0 = x_1(r=0)$ at the steady-state, $\mathbb{C}$ is a constant,
\begin{equation} 
        \mathbb{C} = \left(M_1 - M_2 \right) \frac{{\omega_u}^2}{R_U T}  \; ,
\end{equation}
$R_U$ is the universal gas constant, and $T$ is the temperature. In the EVD-LBM framework, $R_U T \equiv M_1 c_{s,1}^2$. 

Similar to Ref. \cite{Taylor&Krishna}, a binary miscible mixture of uranium isotopes, $\text{U}^{235}\text{F}_6$ (1) and $\text{U}^{238}\text{F}_6$ (2), is contained within the ultracentrifuge, which yields $M_1/M_2= 0.991481$. The constant is set to $\mathbb{C} = -1.2358 \times 10^{-8}$ l.u., which corresponds, for instance, to an ultracentrifuge with a radius of $R = 5$ cm, operating at $T=293.15$ K with $w_u = 1000$ s$^{-1}$. $\tau_1 = 0.6$ and $\text{\DJ}_{12} = 0.08$ are arbitrarily chosen since the steady-state solution is independent of the transport coefficients. An initial average mole fraction $\overline{x_i}$ is specified, from which the corresponding initial mass fraction $\overline{w_{i}} = \overline{x_i} M_i / (\sum_i x_i M_i)$ is determined. The mixture and species densities are then initialized as $\rho = 1$ l.u. and $\rho_i = \overline{w_{i}} \rho$. Both the species and mixture velocities are set to zero. An initial velocity $\mathbf{u}_{i_0} = \mathbf{u}^{eq}_i - (0.5 \mathbf{F}_i \delta t / \rho_i)$ is defined and used in place of the equilibrium velocity in Eq. (\ref{eq:feqi0}) to initialize the distribution functions ($f^i_\alpha = f^{i(0)}_\alpha$) consistently with the applied force field. Finally, the top and bottom walls are handled with periodic conditions, while the boundary scheme proposed in the \ref{sec:bouce_back} is applied to the impermeable solid lateral walls.

\begin{figure}[h]
    	\centering
    	\includegraphics[width=0.65\linewidth]{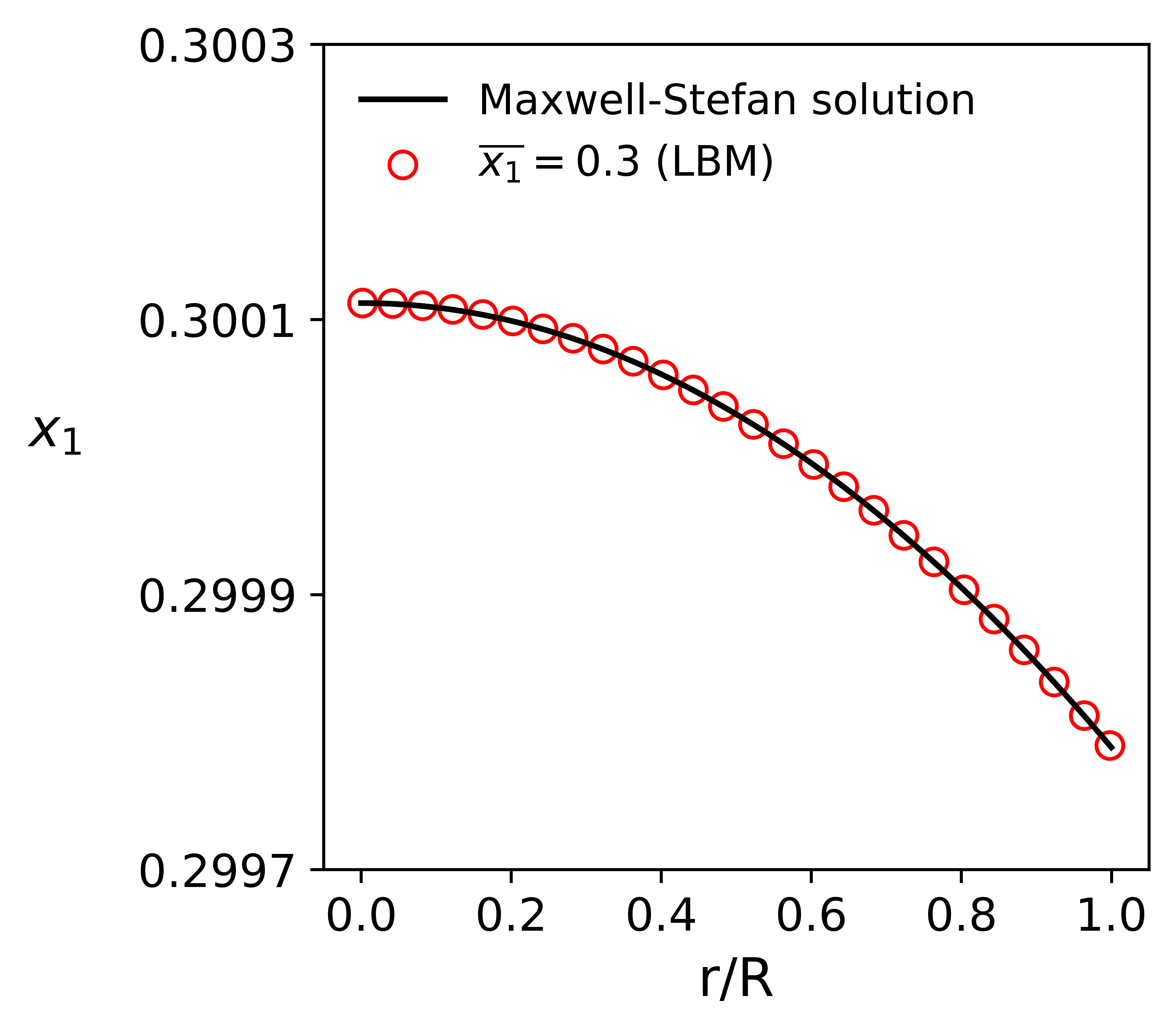}
    	\caption{Concentration profiles for the ultracentrifuge benchmark compared between EVD-LBM simulations and analytical solutions.}    	\label{fig:Ultracentrifuge_concentration_profile}
    \end{figure}

\begin{figure}[H]
    	\centering
    	\includegraphics[width=0.65\linewidth]{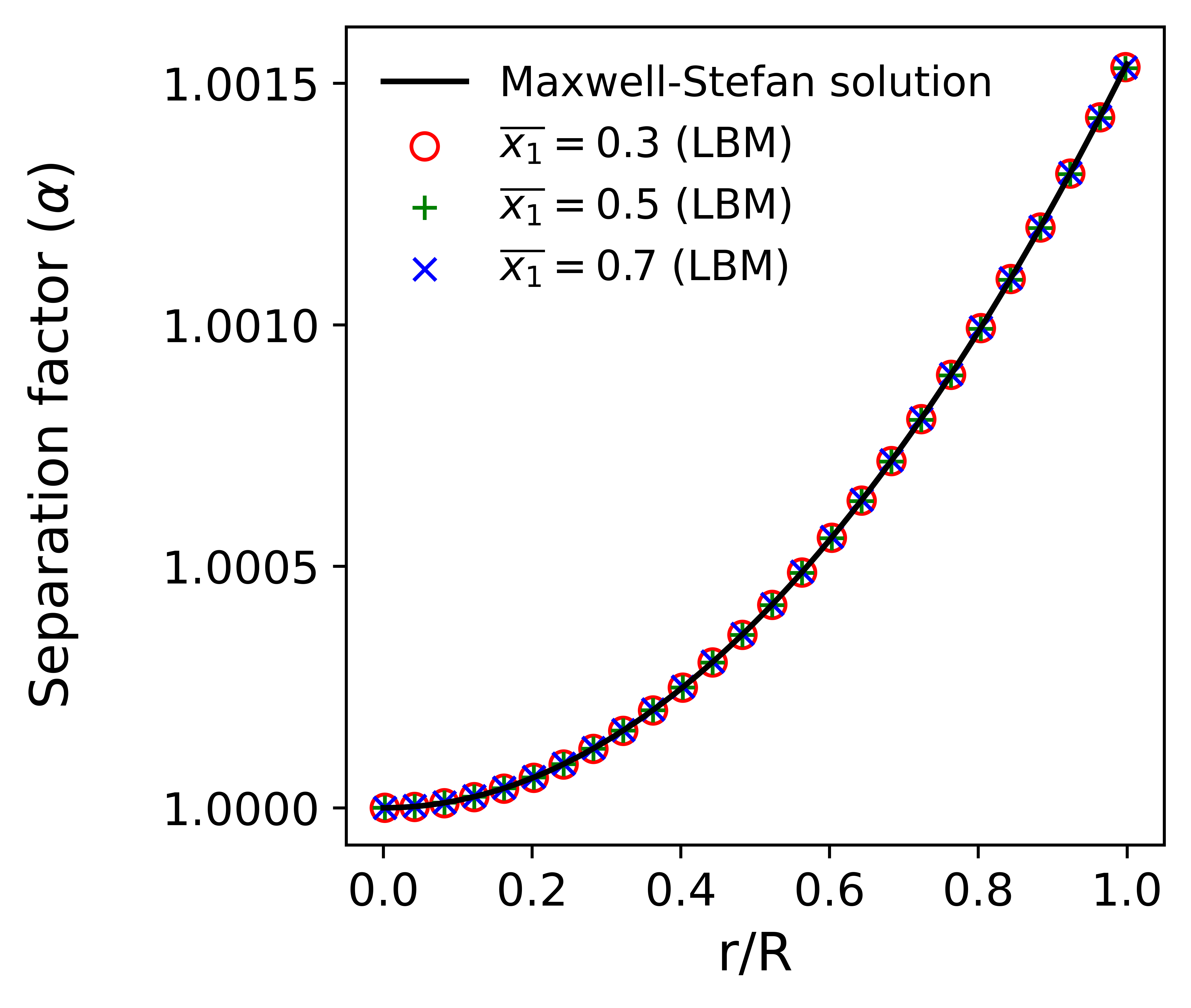}
    	\caption{Variation of the separation factor across the domain for the ultracentrifuge benchmark with three different initial average mole fractions.}    	\label{fig:Ultracentrifuge_alpha_profile}
    \end{figure}

Fig. \ref{fig:Ultracentrifuge_concentration_profile} shows the accurate prediction of the concentration profile for $\overline{x_1}=0.3$ compared to the analytical solution.  This confirms the effectiveness of the forcing approach developed within the EVD model and its reliable physical performance. As expected, the concentration of the lightest species decreases with increasing $r$ as a result of the centrifugal force field acting on the species. We note that the narrow range of the $x_1$ scale arises from the inherent limitations of the ultracentrifuge and the challenge of separating the isotopes. As a result, the separation factor is quite small, as shown in Fig. \ref{fig:Ultracentrifuge_alpha_profile} for cases $\overline{x_1}=0.3$, $\overline{x_1}=0.5$, and $\overline{x_1}=0.7$. The curves for these different cases collapse into a single curve because, as shown in Eq. (\ref{eq:separation_factor_eq}), the separation factor is independent of the mole fraction. Also, according to Eq. (\ref{eq:separation_factor_eq}), an increase in the radius of the ultracentrifuge leads to an increase in the separation factor.

\subsection{Case III: permeable Couette flow (no external force)}\label{sec:Couette}

\begin{figure}[h]
    	\centering
    	\includegraphics[width=0.65\linewidth]{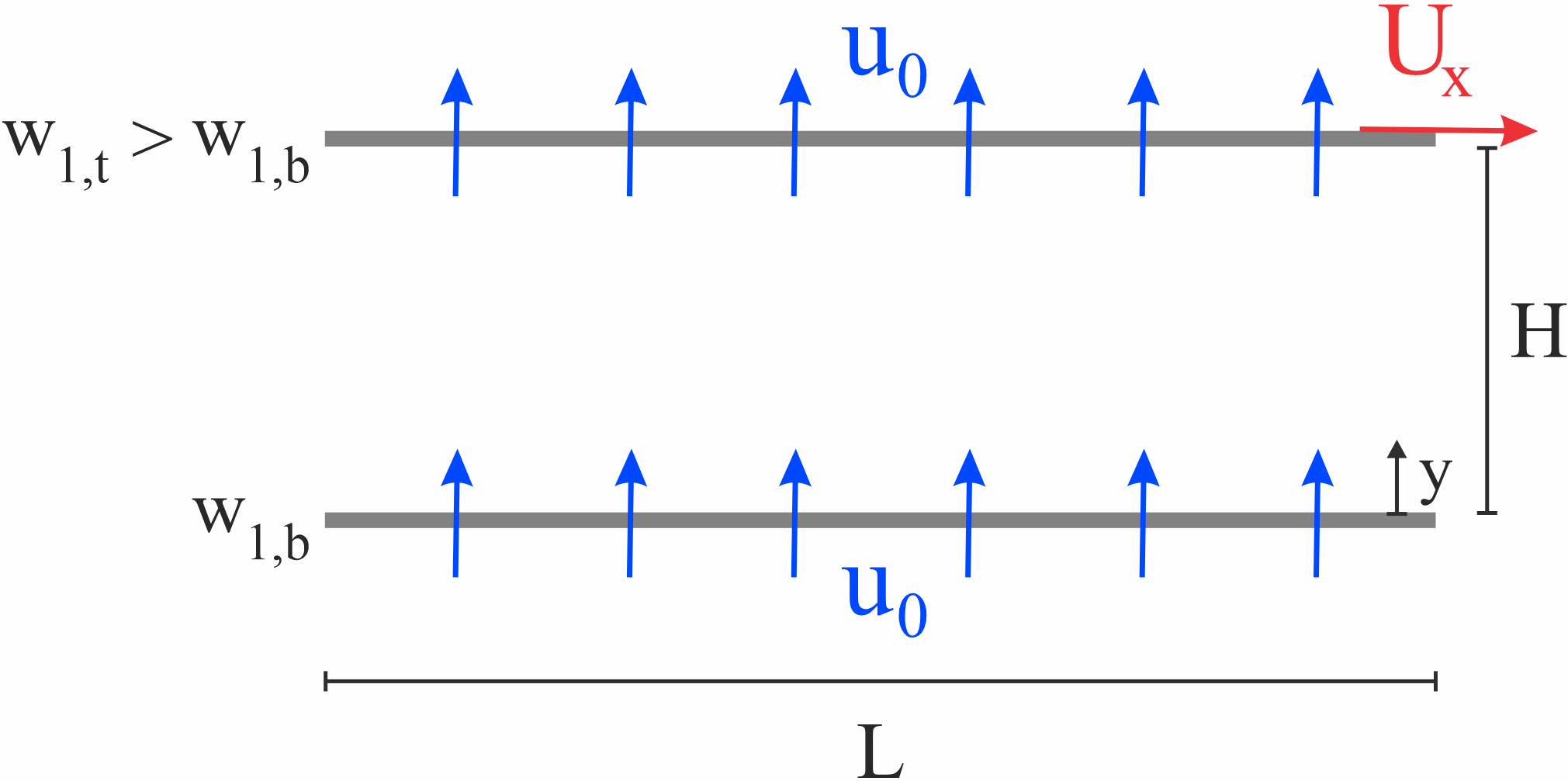}
    	\caption{Sketch of the two-dimensional Couette flow with suction-injection. H and L denote the domain height and length, and $w_{1,t}$ and $w_{1,b}$ represent the mass fractions of an imaginary species 1 at the top and bottom walls.}
    	\label{fig:Sketch_Couette}
    \end{figure}

In Sections \ref{sec:gravitational_force} and \ref{sec:centrifugal_force}, we demonstrated that the proposed forcing approach can handle actual problem simulations where forces acting on species are crucial, a capability yet to be deeply investigated by any existing multifluid EVD-LBM model for mass transfer. In this section, we will demonstrate that the corrective forcing contributions, i.e., the terms $S_{\text{p},\alpha}^i$, $S_{\rho,\alpha}^i$, and $S_{\text{diff},\alpha}^i$ in Eq. (\ref{eq:improved_forcing_scheme}), are associated with Reynolds and Péclet dimensionless numbers. Furthermore, although these terms are necessary for a rigorous recovery of the macroscopic governing equations, they may be numerically irrelevant in certain cases. For this purpose, the Couette flow problem with suction-injection depicted in Fig. \ref{fig:Sketch_Couette} is simulated here. Although no forces act on species in this benchmark, the corrective terms can still emerge. This is because they are incorporated into the LBE through a forcing term, but they are not mathematically tied to the existence of force fields. Instead, they depend on the presence of velocity, density, and concentration fields, all of which are provided by this benchmark. Hence, this benchmark was chosen for its ability to assess the numerical contribution of the corrective terms.

In the permeable Couette flow problem, a binary miscible mixture flows with a velocity $u_0$ between permeable walls separated by a distance $H = 1000$ nodes. Two hypothetical species (1 and 2) are considered to simplify the analysis, assuming $M_2 = 2M_1$ and $\tau_1 = 0.6$. The mass fractions are fixed at the bottom wall ($w_{1,b} = 0.1$) and at the top wall ($w_{1,t} = 0.9$). The velocity $u_0$ and the binary diffusion coefficient $\text{\DJ}_{12}$ are adjusted to account varying Reynolds ($Re = u_0 H / \nu$) and Péclet ($Pe = u_0 H / \text{\DJ}_{12}$) numbers, spanning the range $[1,50,100,150,200]$. The cross-collision parameter $\tau_{12}$ is calculated using Eq. (\ref{eq:Luo&Girimaji_tau}). The top wall moves to the right at a velocity of $U_x = 0.01$ l.u., with the domain length adjusted to $L=3$ nodes and periodic conditions applied at the lateral walls. The mass fractions and the velocities $U_x$ and $u_0$ are imposed on the top and bottom walls employing an extended Zou-He strategy adapted for the EVD model \cite{Tong_etal2014}. The mixture is initialized with zero velocity and density equals $\rho_0 = 1$ l.u., with species densities $\rho_i(y=0) = w_{i,b} \rho_{0}$ and $\rho_i(y\neq0) = w_{i,t} \rho_{0}$. Following the methodology of Tong et al. \cite{Tong_etal2014}, a rescaling factor $\text{RF}(t) = \rho_0 H L / \sum_{\textbf{x}} \rho(\textbf{x},t)$ is applied to normalize the distribution functions every 1000 steps, as the densities at the open boundaries tend to increase over time since they are not fixed. The forcing scheme presented in Eq. (\ref{eq:improved_forcing_scheme}) is intentionally implemented with $\mathbf{F}_i = \mathbf{0}$ to evaluate the relevance of the remaining terms, identified here as (i) the diffusion-driven contribution and (ii) the grouped density and pressure contribution. This segregation into two groups results from the nondimensionalization of the species momentum equation, which reveals that its dimensionless counterpart depends on $Pe \textbf{C}^*_{\textbf{diff},i}$ and $ (\textbf{C}^*_{\mathbf{\rho},i} - \textbf{C}^*_{\textbf{p},i})/Re$, where the asterisk (*) indicates a dimensionless term. Hence, the simulations are performed both with and without each contribution (i and ii). The reader can find the nondimensionalization procedure in the supplemental material.

Fig. \ref{fig:Transient_state_profiles_Re100_Pe20}a shows that the concentration profiles obtained by considering the corrective forcing contributions are very similar to those without any correction, i.e., when all forcing contributions are ignored. This is observed in both unsteady ($t^* < 1$) and steady-state ($t^* = 1$) profiles. One reason for this huge similarity is that the corrective contributions appear only in the species momentum conservation equation, thereby leaving the species and mixture mass conservation equations free of spurious terms. Fig. \ref{fig:Transient_state_profiles_Re100_Pe20}b presents a similar degree of alignment in the velocity profiles. Still, since spurious artifacts arise in the species momentum equation, the explanation for the observed similarity is that these spurious terms are insignificant under the simulated conditions ($Pe = 20$ and $Re = 100$), rendering the use of corrective contributions irrelevant. Consequently, similar velocity fields are generated, leading to equivalent effects in the advection of concentration and further also contributing to the similarity of the concentration profiles.

    \begin{figure}[!h]
    	\centering
    	\includegraphics[width=0.55\linewidth]{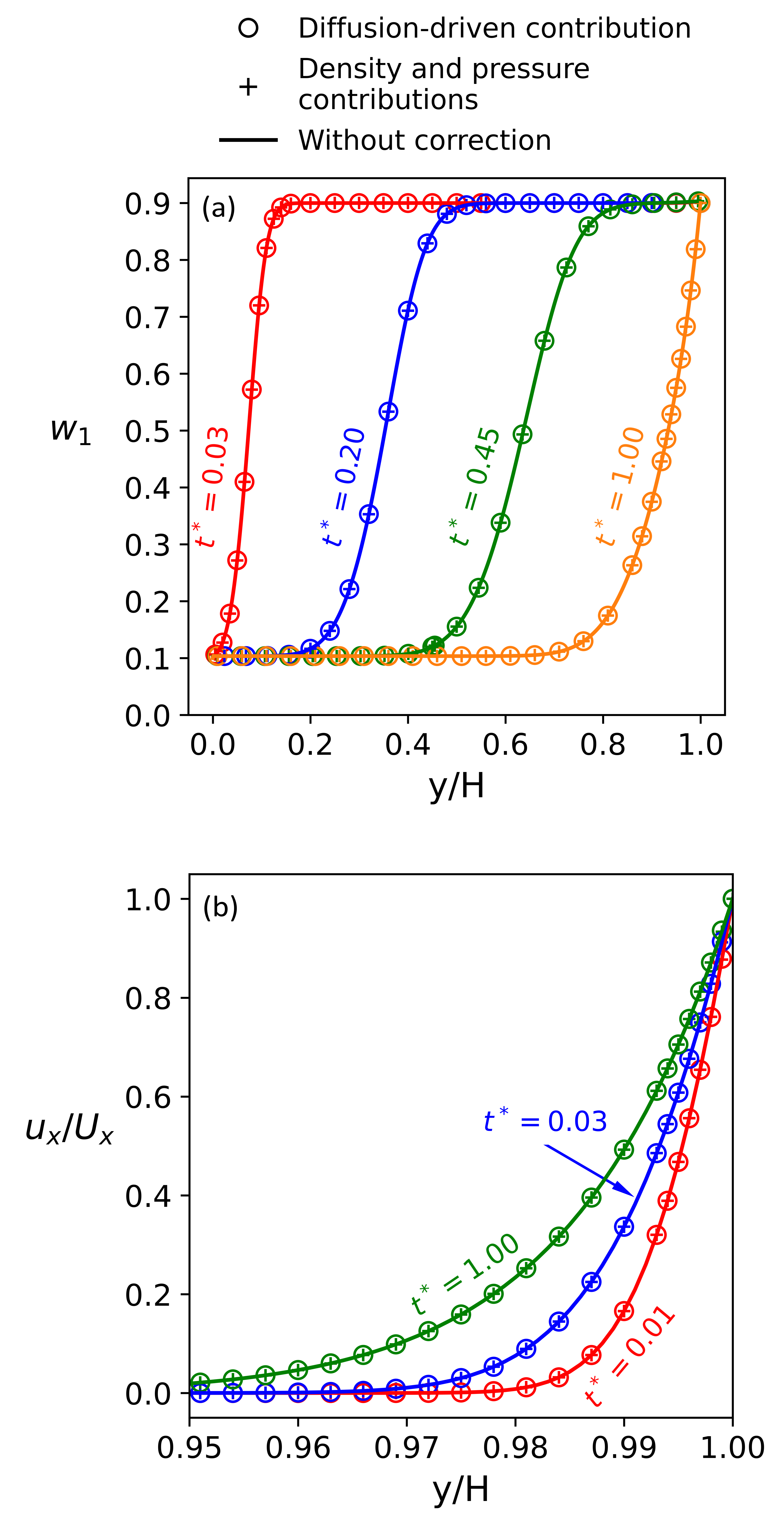}
    	\caption{Evolution of the (a) concentration and (b) velocity profiles over the dimensionless time $t^*$, for $Pe=20$ and $Re = 100$. The solid line represents the numerical simulated results obtained without corrective contributions. The reference time for nondimensionalization is set to 100,000 steps.}    	\label{fig:Transient_state_profiles_Re100_Pe20}
    \end{figure}

\begin{figure}[!h]
    	\centering
    	\includegraphics[width=0.55\linewidth]{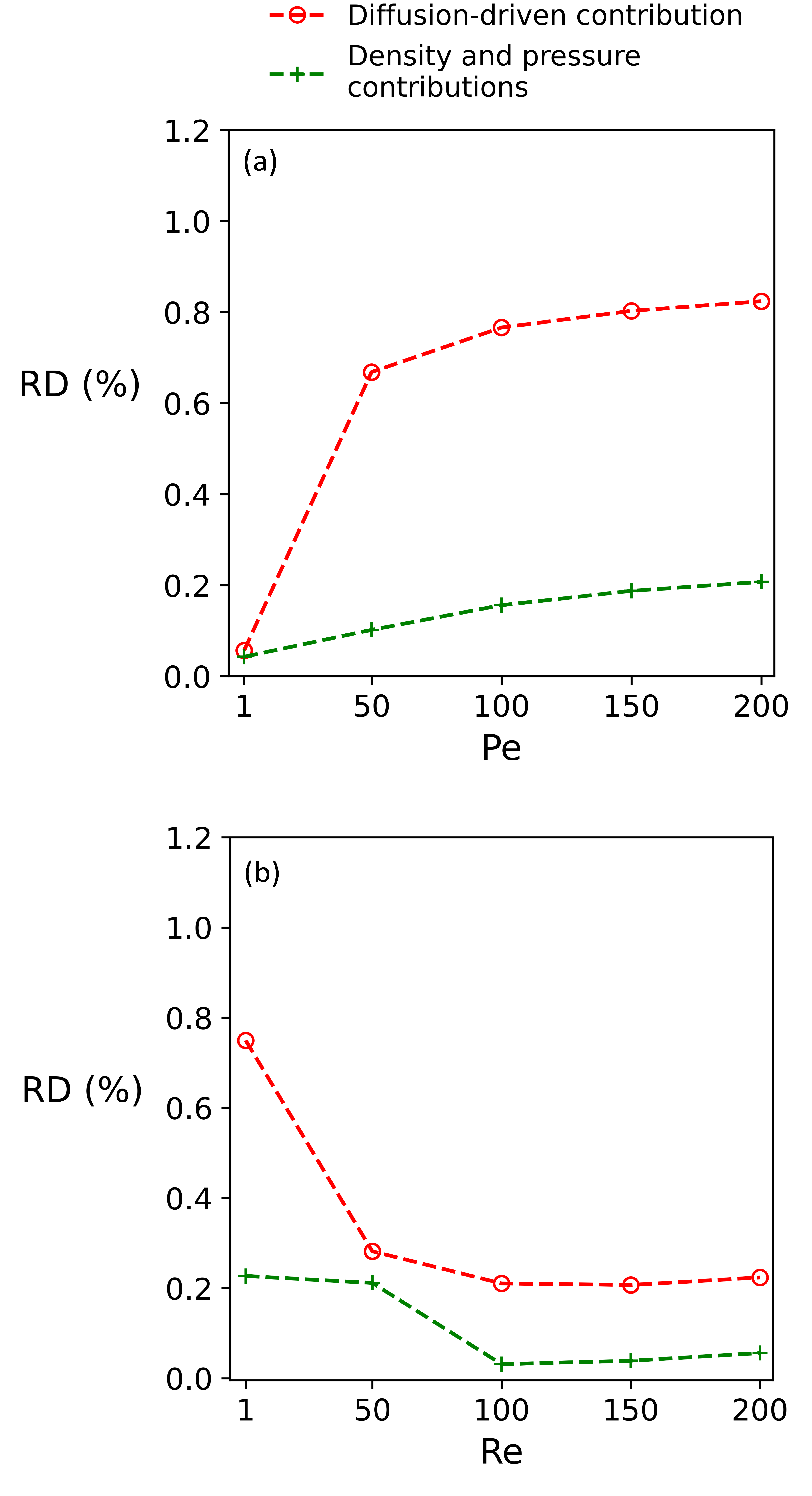}
    	\caption{Maximum relative deviations between the velocity profiles obtained with forcing corrections and those obtained without corrections, for different values of (a) $Pe$ (with a fixed $Re = 20$) and (b) Re (with a fixed $Pe = 20$).}    	\label{fig:RD_related_to_WForce}
    \end{figure}

However, as indicated by the nondimensionalization of the species momentum equation, the diffusion-driven contribution is influenced by $Pe$, while the density and pressure contributions depend on $1/Re$. This implies that the impact of spurious artifacts varies with these dimensionless numbers, suggesting that the corrective contributions could become more significant under certain specific conditions. Particularly, the nondimensionalization indicates that the diffusion-driven contribution becomes more pronounced at higher $Pe$, while the density and pressure contributions become more pronounced at lower $Re$. This behavior is observed in Fig. \ref{fig:RD_related_to_WForce}, which presents the maximum relative deviations (RD) for various $Pe$ and $Re$ conditions. Here, RD is defined by
\begin{equation} 
        \text{RD} (\%) = \frac{ | u_{x,\text{LBM}} - u_{x,\text{WF}} |_{\text{max}}  }{ u_{x,\text{WF}} } \times 100  \; ,
\end{equation}
where $u_{x,\text{LBM}}$ and $u_{x,\text{WF}}$ represent the velocities from the EVD-LBM simulations with and without forcing contributions. In addition to this behavior, Fig. \ref{fig:RD_related_to_WForce} also reports that RD increases with $Pe$ for simulations including the density and pressure contributions, and decreases with $Re$ (especially for $Re<100$) for those incorporating the diffusion-driven contributions, a result not theoretically predicted by the nondimensionalization. This behavior remains unclear to us and requires further investigation to fully understand the underlying mechanisms at play.

The diffusion-driven contribution has a greater impact compared to the density and pressure contributions. However, all simulations achieve modest deviations (RD $<1\%$), even under the extreme values of the investigated range of $Pe$ and $Re$. Hence, although the corrective contributions may affect the velocity profiles, their impact is negligible compared to the methodology without any corrections, corroborating that the corrective contributions can be omitted without significant numerical loss. Despite being an acceptable simplification, we emphasize that the corrective terms derived in this work remain necessary for a rigorous recovery of the macroscopic momentum equations.

\begin{figure}[!h]
    	\centering
    	\includegraphics[width=0.55\linewidth]{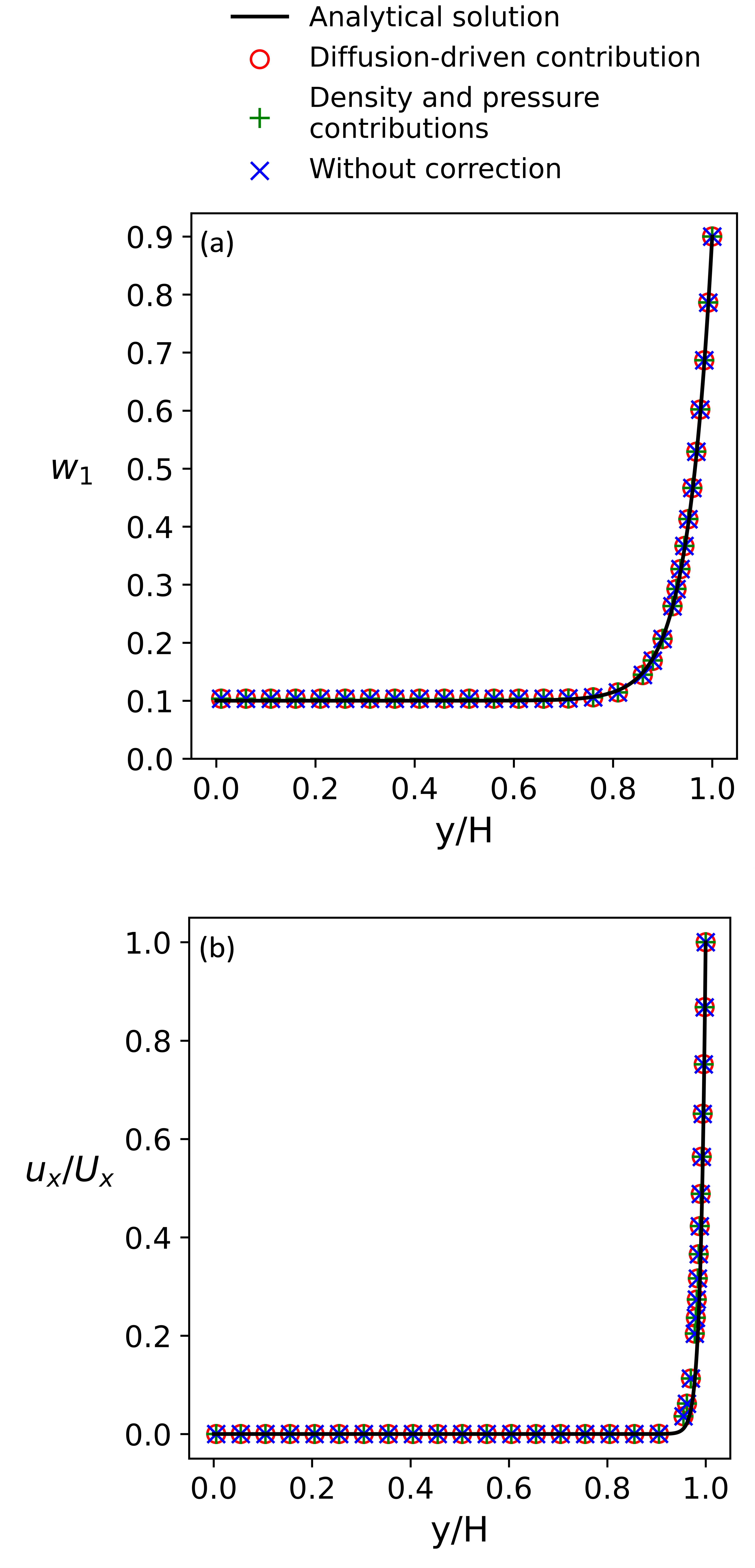}
    	\caption{Steady-state (a) concentration and (b) velocity profiles of the permeable Couette flow for $Pe = 20$ and $Re = 100$. The term ``without correction'' indicates that no corrective contributions are applied.}    	\label{fig:Steady_state_profiles_Re100_Pe20}
    \end{figure}

As a final point, the steady-state numerical profiles can be evaluated against the analytical solutions of the permeable Couette flow benchmark provided by \cite{Zhang_etal2012}
\begin{equation} 
        \frac{w_1(y) - w_{1,b}}{w_{1,t} - w_{1,b}} = \frac{ \exp\left(\frac{Pe \; y}{H}\right) - 1}{\exp(Pe) - 1}  \; ,
\end{equation}
\begin{equation} 
        \frac{u_x(y)}{U_x} = \frac{ \exp\left(\frac{Re \; y}{H}\right) - 1}{\exp(Re) - 1}  \; .
\end{equation}
The simulations yield profiles that are not only comparable to each other but also closely match the analytical solutions, as illustrated in Fig. \ref{fig:Steady_state_profiles_Re100_Pe20} for $Pe=20$ and $Re=100$.

\section{Conclusion}\label{sec:conclusion}

Here, we proposed a forcing approach within the explicit velocity-difference LBM framework. By performing the Chapman-Enskog expansion, the mesoscopic modeling recovers the macroscopic species and mixture mass conservation equations, the Navier-Stokes equation with external forcing term, and the full Maxwell-Stefan equation for ideal miscible mixtures at low Knudsen numbers. We considered gravitational and centrifugal forces acting on species as benchmarks to establish the physical consistency and applicability of the proposed forcing approach in systems where external forces are crucial for realistic and accurate mass transfer simulations. The proposed forcing approach effectively accounts for significant forcing effects and aligns closely with expected analytical solutions when available. Notably, we demonstrated that the corrective forcing contributions derived through the Chapman-Enskog analysis, necessary for mathematically eliminating spurious artifacts in the species momentum equation, are associated with the Reynolds and Péclet numbers. Depending on the simulated case, these contributions may have minimal impact on numerical results, thus allowing for their omission when appropriate.

We also proposed and implemented a boundary scheme for impermeable solid walls to enable these accurate simulations with spatial interpolations, ensuring proper mass conservation while enforcing the wall velocity on the fluid adjacent to the wall. Ultimately, the complete methodology investigated in this work effectively addresses forcing effects that previous EVD-LBM models could not address, while maintaining a multifluid perspective rather than the single-fluid approach used in passive scalar models. The proposed forcing approach creates future opportunities to explore the integration of complex multiphase effects and non-ideal behaviors into the multifluid mass transfer modeling via carefully defined force interactions, including scenarios involving imposed force fields or interaction forces between species and surfaces. 

\section*{Acknowledgments}
We gratefully acknowledge CNPq (Conselho Nacional de Desenvolvimento Científico e Tecnológico), FAPERJ (Fundação Carlos Chagas Filho de Amparo à Pesquisa do Estado do Rio de Janeiro), Shell, and ANP (Agência Nacional do Petróleo) for the financial support.

\appendix
\section{A boundary scheme for impermeable walls within the EVD model} \label{sec:bouce_back}

The halfway bounce-back scheme is a well-known approach in LBM for modeling no-slip solid boundaries, ensuring that the fluid velocity adjacent to the boundary mirrors the wall velocity. For stationary walls, it is mathematically expressed by
\begin{equation} \label{eq:standard_BBscheme}
        f_{\overline{\alpha}}(\textbf{x}, t + \delta t) = f_{\alpha} (\textbf{x}, t) \; ,
\end{equation}
where $\overline{\alpha}$ is the opposite direction of $\alpha$ \cite{Ladd_1994, Ginzbourg&Adler1994}. 

In the mass transfer framework using the EVD modeling, bounce-back rules have been employed to model impermeable stationary walls typically by storing distribution values in virtual lattice nodes (fullway bounce-back scheme) \cite{Tong_etal2014,Ma&Chen2015}. Instead, the halfway bounce-back scheme is generally preferred in LBM to avoid managing a larger simulated domain and increasing computational memory. For setups where species have the same molecular mass, the standard halfway bounce-back scheme in Eq. (\ref{eq:standard_BBscheme}) not only aligns the fluid behavior near the boundary with the expected wall velocity but also ensures that there is no mass flux through impermeable walls. However, as will be demonstrated, this scheme is incapable of ensuring mass conservation when spatial interpolations are required. In the following, we propose modifications to the standard halfway bounce-back rule to ensure accurate mass conservation at impermeable solid boundaries when considering species with different molecular masses.

\begin{figure}[h]
    	\centering
    	\includegraphics[width=0.45\linewidth]{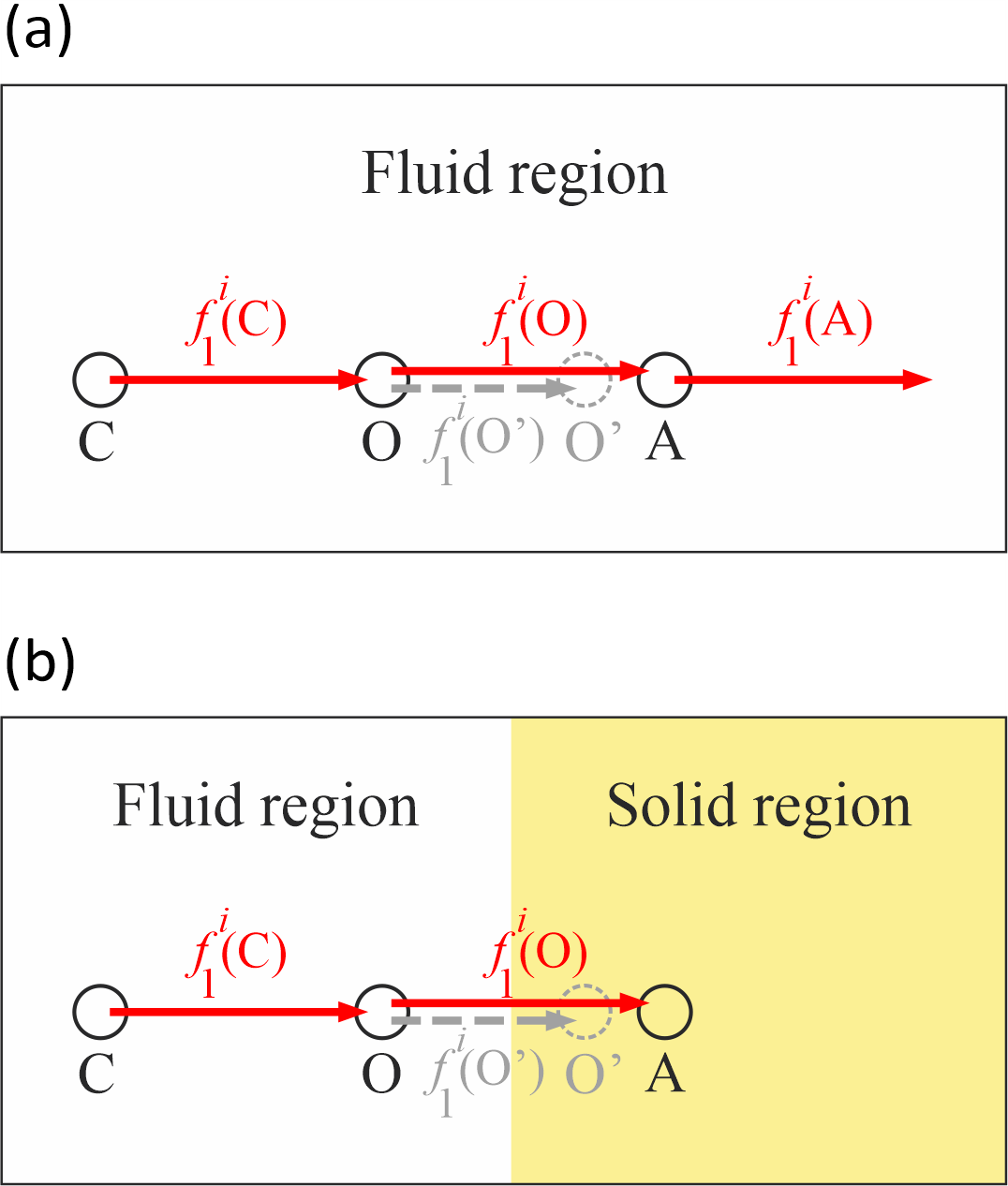}
    	\caption{The streamed distribution $f_1^i(O')$ is redistributed during spatial interpolation to compute $f_1^i(O)$, $f_1^i(A)$, and $f_1^i(C)$ in (a), and $f_1^i(O)$ and $f_1^i(C)$ in (b).  Solid black circles, denoted by pure letters, mark the discrete lattice points, while dashed light-grey circles, indicated by single-primed letters, represent the off-lattice positions to which species $i \neq 1$ have moved.}
    	\label{fig:Reformulated_BB_alfa_orthogonal}
    \end{figure}

The conservation of $f_\alpha^i$ is rigorously maintained in the nodes far from the boundaries, where streaming and spatial interpolations occur without restrictions. This is confirmed by inspecting if all information from the streamed distribution $f_\alpha^i(X')$, for a specific $\alpha$ and location $X'$, is fully utilized without any loss or gain in the interpolation scheme. For example, for $\alpha = 1$ and $X'=O'$, the information contained in the streamed distribution $f_1^i(O')$ is redistributed during spatial interpolation to compute $f_1^i(O)$, $f_1^i(A)$, and $f_1^i(C)$, as illustrated in Fig. \ref{fig:Reformulated_BB_alfa_orthogonal}a. Based on Eq. (\ref{eq:Second_order_spatial_interpolation}), they are calculated by
\begin{equation}  
    \label{eq:alpha_orthogonal_BB_reformulated}
    \begin{aligned} 
    f_1^i(O) &= (1-\xi_1^2) f_1^i(O') + [...] \; ,  \\
    f_1^i(A) &= \frac{\xi_1 (1+\xi_1) }{2} f_1^i(O') + [...] \; ,  \\
    f_1^i(C) &= - \frac{\xi_1 (1-\xi_1) }{2} f_1^i(O') + [...] \; ,
    \end{aligned}
\end{equation}
where $\eta_1 = 0$ is used. To verify whether all information pertaining to $f_1^i(O')$ has been utilized without any loss or gain, the sum of the terms associated with $f_1^i(O')$ must equal 1, which indeed is the case. Hence, $f_1^i(O')$ is conserved. Similarly, $f_\alpha^i(X')$ is conserved in the nodes far from the boundaries for any $\alpha$ orthogonal to the domain axes, i.e., $\alpha=[1,2,3,4]$. A similar analysis can be conducted for non-orthogonal $\alpha$, i.e., $\alpha=[5,6,7,8]$, to observe the conservation of $f_\alpha^i(X')$. Also note that for $\alpha = 0$, the conservation of $f_0^i(X')$ is rapidly verified, as its value is entirely transferred to $f_0^i(X)$, as observed in Eq. (\ref{eq:Second_order_spatial_interpolation}) with $\xi_0 = \eta_0 = 0$.

However, the conservation of $f_\alpha^i(X')$ is not guaranteed in the nodes near impermeable boundaries if the standard bounce-back scheme is employed, as the one represented in Fig. \ref{fig:Reformulated_BB_alfa_orthogonal}b. Proceeding with the case in which $\alpha$ is orthogonal to the axes (say $\alpha = 1$), no distribution function leaves node A in Fig. \ref{fig:Reformulated_BB_alfa_orthogonal}b because it is a solid node. Hence, $f_1^i(O')$ is redistributed during spatial interpolation to compute $f_1^i(O)$ and $f_1^i(C)$ only,
\begin{equation}  
    \label{eq:alpha_orthogonal_near_wall_BB_reformulated}
    \begin{aligned} 
    f_1^i(O) &= (1-\xi_1^2) f_1^i(O') + [...] \; , \\
    f_1^i(C) &= - \frac{\xi_1 (1-\xi_1) }{2} f_1^i(O') + [...]  \; .
    \end{aligned}
\end{equation}

The comparison between Eqs.  (\ref{eq:alpha_orthogonal_BB_reformulated}) and (\ref{eq:alpha_orthogonal_near_wall_BB_reformulated}) reveals that the term $\xi_1 (1+\xi_1) f_1^i(O') /2$ is missing in Eq. (\ref{eq:alpha_orthogonal_near_wall_BB_reformulated}) for calculating any distribution function, leaving the conservation of $f_1^i(O')$ unresolved near the impermeable boundary. Similar to this unused term, a portion of $f_3^i(O')$, namely $\xi_3 (1+\xi_3) f_3^i(O') /2$, was not employed to compute $f_1^i(A)$ either. Consequently, these terms will become contributions to calculate $f_{3}^i(O)$ (where $\overline{\alpha}=3$ in this case). Hence, the scheme that we propose for solid and impermeable boundaries on right walls reads   
\begin{equation} \label{eq:reformulated_bb_f3}
    \begin{aligned}
        f_3^i (O) = (1-\xi_3^2) f_3^i(O') + \frac{\xi_3 (1+\xi_3)}{2} f_3^i(C') + \frac{\xi_1 (1+\xi_1)}{2} f_1^i(O')  \\ + \frac{\xi_3 (1+\xi_3)}{2} f_3^i(O')    \; ,
    \end{aligned}
\end{equation}
 where the first two terms on the right-hand side originally appear in the calculation of $f_3^i(O)$ from Eq. (\ref{eq:Second_order_spatial_interpolation}), and the last two terms are the contributions that would be used to calculate $f_1^i(A)$ and $f_3^i(A)$, respectively. Substituting $\xi_3=-\xi_1$ and rewriting Eq. (\ref{eq:reformulated_bb_f3}) in a more convenient format yields
\begin{equation}
    \begin{aligned}
        f_3^i (O) = f_3^i(O') + \frac{\xi_1^2}{2}\left[f_1^i(O') + f_3^i(C') - f_3^i(O') \right] \\ + \frac{\xi_1}{2}\left[f_1^i(O') - f_3^i(C') - f_3^i(O') \right]   \; .
    \end{aligned}
\end{equation}

Likewise, the previous methodology can be applied to non-orthogonal directions. The proposed scheme for the diagonal orientation for solid impermeable boundaries on right walls is obtained as
\begin{equation} 
    \begin{aligned}
        f_\alpha^i (O) = \left[ (1-\eta_\alpha^2) (1-\xi_\alpha^2) + \frac{\xi_\alpha (1+\xi_\alpha)}{2} \right] f_\alpha^i(O') \\ + \frac{\xi_\alpha (1 + \xi_\alpha) (1 - \eta_\alpha^2)}{2}  f_\alpha^i(C') + \frac{\eta_\alpha (1 + \eta_\alpha) (1 - \xi_\alpha^2)}{2}  f_\alpha^i(D') \\ + \frac{\eta_\alpha \xi_\alpha (1 + \eta_\alpha) (1 + \xi_\alpha)}{4}  f_\alpha^i(G') - \frac{\eta_\alpha (1 - \eta_\alpha) (1 - \xi_\alpha^2)}{2}  f_\alpha^i(B') \\ - \frac{\eta_\alpha \xi_\alpha (1 + \xi_\alpha) (1 - \eta_\alpha)}{4}  f_\alpha^i(F') - \frac{\xi_\alpha (1 - \xi_\alpha)}{2}  f_{\overline{\alpha}}^i(O')\; ,
    \end{aligned}
\end{equation}
where $\alpha = 6$ or $\alpha = 7$.

Below, the reader will find the final proposed scheme for the remaining three walls of a two-dimensional domain using the D2Q9 lattice arrangement:

\begin{itemize}
    \item solid impermeable boundaries on left walls:
\end{itemize}
\begin{equation} 
    \begin{aligned}
        f_1^i (O) = f_1^i(O') + \frac{\xi_1^2}{2}\left[f_3^i(O') + f_1^i(A') - f_1^i(O') \right] \\ + \frac{\xi_1}{2}\left[f_3^i(O') - f_1^i(A') - f_1^i(O') \right]   \; ,
    \end{aligned}
\end{equation}
\begin{equation} 
    \begin{aligned}
        f_\alpha^i (O) = \left[ (1-\eta_\alpha^2) (1-\xi_\alpha^2) - \frac{\xi_\alpha (1-\xi_\alpha)}{2} \right] f_\alpha^i(O') \\ - \frac{\xi_\alpha (1 - \xi_\alpha) (1 - \eta_\alpha^2)}{2}  f_\alpha^i(A') - \frac{\eta_\alpha (1 - \eta_\alpha) (1 - \xi_\alpha^2)}{2}  f_\alpha^i(B') \\ + \frac{\eta_\alpha (1 + \eta_\alpha) (1 - \xi_\alpha^2)}{2}  f_\alpha^i(D') - \frac{\eta_\alpha \xi_\alpha (1 - \xi_\alpha) (1 + \eta_\alpha)}{4} f_\alpha^i(H') \\ + \frac{\eta_\alpha \xi_\alpha (1 - \xi_\alpha) (1 - \eta_\alpha)}{4}  f_\alpha^i(E') + \frac{\xi_\alpha (1 + \xi_\alpha)}{2}  f_{\overline{\alpha}}^i(O')\; ,
    \end{aligned}
\end{equation}
where $\alpha = 5$ or $\alpha = 8$;

\begin{itemize}
    \item solid impermeable boundaries on upper walls:
\end{itemize}
\begin{equation} 
    \begin{aligned}
        f_4^i (O) = f_4^i(O') + \frac{\eta_2^2}{2}\left[f_2^i(O') + f_4^i(D') - f_4^i(O') \right] \\ + \frac{\eta_2}{2}\left[f_2^i(O') - f_4^i(D') - f_4^i(O') \right]   \; ,
    \end{aligned}
\end{equation}
\begin{equation} 
    \begin{aligned}
        f_\alpha^i (O) = \left[ (1-\eta_\alpha^2) (1-\xi_\alpha^2) + \frac{\eta_\alpha (1+\eta_\alpha)}{2} \right] f_\alpha^i(O') \\ + \frac{\eta_\alpha (1 - \xi_\alpha^2) (1 + \eta_\alpha)}{2}  f_\alpha^i(D') + \frac{\xi_\alpha (1 + \xi_\alpha) (1 - \eta_\alpha^2)}{2}  f_\alpha^i(C') \\ - \frac{\xi_\alpha (1 - \xi_\alpha) (1 - \eta_\alpha^2)}{2}  f_\alpha^i(A') + \frac{\eta_\alpha \xi_\alpha (1 + \eta_\alpha) (1 + \xi_\alpha)}{4}  f_\alpha^i(G') \\ - \frac{\eta_\alpha \xi_\alpha (1 - \xi_\alpha) (1 + \eta_\alpha)}{4}  f_\alpha^i(H') - \frac{\eta_\alpha (1 - \eta_\alpha)}{2}  f_{\overline{\alpha}}^i(O')\; ,
    \end{aligned}
\end{equation}
where $\alpha = 7$ or $\alpha = 8$;

\begin{itemize}
    \item solid impermeable boundaries on bottom walls:
\end{itemize}
\begin{equation} 
    \begin{aligned}
        f_2^i (O) = f_2^i(O') + \frac{\eta_2^2}{2}\left[f_4^i(O') + f_2^i(B') - f_2^i(O') \right] \\ + \frac{\eta_2}{2}\left[f_4^i(O') - f_2^i(B') - f_2^i(O') \right]   \; ,
    \end{aligned}
\end{equation}
\begin{equation} 
    \begin{aligned}
        f_\alpha^i (O) = \left[ (1-\eta_\alpha^2) (1-\xi_\alpha^2) - \frac{\eta_\alpha (1-\eta_\alpha)}{2} \right] f_\alpha^i(O') \\ + \frac{\xi_\alpha (1 - \eta_\alpha^2) (1 + \xi_\alpha)}{2}  f_\alpha^i(C') - \frac{\xi_\alpha (1 - \xi_\alpha) (1 - \eta_\alpha^2)}{2}  f_\alpha^i(A') \\ - \frac{\eta_\alpha (1 - \eta_\alpha) (1 - \xi_\alpha^2)}{2}  f_\alpha^i(B') - \frac{\eta_\alpha \xi_\alpha (1 + \xi_\alpha) (1 - \eta_\alpha)}{4}  f_\alpha^i(F') \\ + \frac{\eta_\alpha \xi_\alpha (1 - \xi_\alpha) (1 - \eta_\alpha)}{4}  f_\alpha^i(E') + \frac{\eta_\alpha (1 + \eta_\alpha)}{2}  f_{\overline{\alpha}}^i(O')\; ,
    \end{aligned}
\end{equation}
where $\alpha = 5$ or $\alpha = 6$.

Noteworthily, the DLS approach and spatial interpolations become unnecessary when all species have the same molecular mass. This results in $|\xi_\alpha| = 1$ and $|\eta_\alpha| = 1$ for $\alpha \neq 0$, simplifying the boundary scheme proposed here into the standard halfway bounce-back shown in Eq. (\ref{eq:standard_BBscheme}). For non-stationary walls moving with velocity magnitude $u_w$ aligned with its length (say x-direction), this work suggests taking into account the term $-2 \omega_\alpha \rho_i w_i (e_\alpha^i|_x u_w)/ c_{s,i}^2$. For pure species ($\rho_i = \rho$ and $w_i = 1$), this term simplifies to the well-known term used in the standard bounce-back scheme for moving walls, $-2 \omega_\alpha \rho (e_\alpha|_x u_w)/ c_{s}^2$ \cite{Ladd_1994}.

\clearpage
\section*{Nomenclature}

\begin{center}
\begin{longtable}{ll}
$c_i$ & Dimensionless thermal speed of species $i$ \\
$c_{s,i}$ & Speed of sound for species $i$ \\
${\text{\DJ}}_{ij}$ & Diffusion coefficient for the $i$-$j$ pair \\
$\mathbf{e}_\alpha^i$ & Dimensionless discrete velocity of species $i$ \\
$f^i_\alpha$ & Probability distribution function of species $i$ \\
$f^{i(0)}_\alpha$ & Equilibrium distribution function of species $i$ \\ 
$f^{i(eq)}_\alpha$ & Standard equilibrium distribution function of species $i$ \\ 
$f_\alpha^{i,\text{(neq)}}$ & Non-equilibrium distribution function of species $i$ \\
$\mathbf{F}$ & External total force \\
$\mathbf{F}_i$ & External force acting on species $i$ \\
$g$ & Gravity \\
$Ga$ & Galilei number \\
$H$ & Domain height \\
$\mathbf{j}_i$ & Mass diffusive flux \\
$\mathbf{k}_i$ & Specific force acting on species $i$ \\
$L$ & Domain length \\
$Ma$ & Mach number \\
$M_i$ & Molecular mass of species $i$ \\
$n$ & Total mole number \\
$p$ & Total pressure \\ 
$Pe$ & Péclet number \\
$p_i$ & Partial pressure of species $i$ \\
$R$ & Radius \\
RD & Relative deviation \\
$Re$ & Reynolds number \\
$R_U$ & Universal gas constant \\
$S_\alpha^i$ & Forcing term in the LBE \\
$S_{\text{diff},\alpha}^i$ & Diffusion-driven forcing contribution \\
$S_{\text{p},\alpha}^i$ & Pressure forcing contribution \\
$S_{\rho,\alpha}^i$ & Density forcing contribution \\
$t^*$ & Time required for a system to reach a steady-state concentration \\
$T$ & Temperature \\
$\mathbf{u}$ & Mixture velocity \\
$\mathbf{u}_i^{eq}$ & Equilibrium velocity of species $i$ \\
$U_x$ & Wall velocity in the x-direction \\
$w_i$ & Mass fraction of species $i$ \\
$\overline{w_i}$ & Average mass fraction of species $i$ \\
$w_{1,b}$ & Mass fraction at bottom wall for species 1 \\
$w_{1,t}$ & Mass fraction at top wall for species 1 \\
$x_i$ & Mole fraction \\
$\overline{x_i}$ & Average mole fraction of species $i$ \\
$\alpha(r)$ & Separation factor \\
$\varepsilon$ & Small parameter in the Chapman-Enskog expansion \\
$\Omega^{ij}$ & Collision term for particles $i$ and $j$ \\
$\nu$ & Mixture kinematic viscosity \\ 
$\rho$ & Mixture density \\
$\rho_i$ & Density of species $i$ \\
$\tau_i$ & Relaxation time for self-collisions \\
$\tau_{ij}$ & Relaxation time for cross-collisions \\
$\omega_\alpha$ & Weight \\
$\omega_u$ & Angular velocity \\
\end{longtable}
\end{center}

\section*{Acronyms}

\begin{center}
\begin{tabular}{ll}
BTE & Boltzmann transport equation \\
C-E & Chapman-Enskog \\
DLS & Different lattice speeds \\
EVD & Explicit velocity-difference \\
LBE & Lattice Boltzmann equation \\
LBM & Lattice Boltzmann method \\
M-S & Maxwell-Stefan \\
NSE & Navier-Stokes equation \\
\end{tabular}
\end{center}

\bibliographystyle{elsarticle-num-names}

\end{document}